\documentclass[11pt,a4paper]{article}
\pdfoutput=1

\usepackage{jheppub}
\usepackage{amsfonts,amsmath,bm}
\usepackage{array}
\usepackage{graphicx}

\makeatletter
\def\@fpheader{\relax}
\makeatother

\newcommand\be{\begin{equation}}
\newcommand\ee{\end{equation}}
\newcommand\bea{\begin{eqnarray}}
\newcommand\eea{\end{eqnarray}}
\newcommand\eref[1]{(\ref{#1})}

\newcommand\comment[1]{}

\title{\centering Topological Effects in Neural Network Field Theory}

\author[*]{Christian Ferko,}
\author[*]{James Halverson,}
\author[\dagger,*]{Vishnu Jejjala,}
\author[\ddagger]{Brandon Robinson}

\affiliation[*]{The NSF Institute for Artificial Intelligence and Fundamental Interactions (IAIFI)\\
and Department of Physics, Northeastern University, Boston, MA 02115, USA}

\affiliation[\dagger]{Mandelstam Institute for Theoretical Physics, School of Physics, and NITheCS,\\
University of the Witwatersrand, Johannesburg, WITS 2050, South Africa}

\affiliation[\ddagger]{Institute of Physics, University of Amsterdam, Science Park 904,\\ 1098 XH Amsterdam, Netherlands}

\emailAdd{c.ferko@northeastern.edu}
\emailAdd{j.halverson@northeastern.edu}
\emailAdd{v.jejjala@wits.ac.za}
\emailAdd{b.j.robinson@uva.nl}

\abstract{%
Neural network field theory formulates field theory as a statistical ensemble of fields defined by a network architecture and a density on its parameters.
We extend the construction to topological settings via the inclusion of discrete parameters that label the topological quantum number.
We recover the Berezinskii--Kosterlitz--Thouless transition, including the spin-wave critical line and the proliferation of vortices at high temperatures.
We also verify the T-duality of the bosonic string, showing invariance under the exchange of momentum and winding on $S^1$, the transformation of the sigma model couplings according to the Buscher rules on constant toroidal backgrounds, the enhancement of the current algebra at self-dual radius, and non-geometric T-fold transition functions.
}

\begin{document}
\maketitle
\flushbottom

\section{Introduction}\label{sec:intro}

Neural network field theory (NN-FT) is the idea that a field theory can be formulated~\cite{Halverson:2020trp,Halverson:2021aot} from first principles via the definition of a neural network architecture and a density on its parameters, rather than only as a path integral over spacetime fields. For scalar field theories defined as densities over tempered Schwartz distributions --- the setting of much of constructive field theory --- the NN-FT approach was recently shown to be universal~\cite{Ferko:2026axm}.

In this paper we focus on scalar theories with architecture $\phi_\theta$ with parameters $\theta$.
We will sometimes refer to these network parameters as latent variables.
Inputting a position $x$ provides the value of the field, $\phi_\theta(x) \simeq \phi(x)$.  Equipped with the parameter density, correlation functions may be computed from the partition function 
\be
Z[J] = \mathbb{E}[e^{\int d^dx J(x) \phi(x)}] = \int d\theta\; P(\theta)\; e^{\int d^dx\, J(x) \phi(x)} ~,
\ee
where the second equality defines the parameter space description of the NN-FT. Many architectures have a parameter $N\in \mathbb{N}$ such that in the large-$N$ limit the networks converge to Gaussian processes~\cite{neal,williams,Matthews2018GaussianPB,schoenholz2017correspondence,yangTP1}, which are defined by Gaussian densities over functions and therefore realize free field theories. Interactions may be introduced~\cite{PhysRevE.104.064301,Halverson:2021aot,Demirtas:2023fir,Ferko:2026axm} by deviating from this limit or breaking statistical independence of network parameters.

The philosophy is therefore not merely to use neural networks as flexible variational ans\"atze for approximating a pre-existing theory, but to treat architecture and parameter densities as data that define the theory itself, so that locality~\cite{Demirtas:2023fir}, symmetry~\cite{Maiti:2021fpy}, operator content, and phase structure are encoded directly in ensembles over network parameters.
In this way, NN-FT provides both a conceptual bridge between modern machine learning and field theory and a constructive framework in which increasingly rich systems, including conformal theories~\cite{Halverson:2024axc,Robinson:2025ybg,Capuozzo:2025ozt}, quantum mechanics~\cite{Ferko:2025ogz}, fermionic and supersymmetric theories~\cite{Frank:2025zuk,Huang:2025ipy}, and worldsheet string theory~\cite{Frank:2026bui}, can be realized and studied through statistics of neural parameter space and architectures that push them forward to function space.

The present paper introduces topological effects into NN-FT. Consistent with the spirit that a NN-FT is defined by an architecture and its parameters, we realize topological sectors via the introduction of discrete network parameters that realize topological quanta.

The question is especially sharp for compact bosons.
A non-compact Gaussian field describes smooth local fluctuations, but compactness introduces data that are discrete from the start: winding numbers, momentum sectors, and defect sectors.
In ordinary path integrals, these ingredients are not produced automatically by a Gaussian fluctuation integral; they enter through the choice of field space and measure.
The same issue therefore arises in NN-FT: a single-valued Gaussian sampler can represent the local sector cleanly, yet by itself it does not generate the full set of topological sectors needed for the compact theory.
The NN-FT constructions we propose mirror the logic of the path integral by enlarging parameter space to include both continuous and discrete parameters.

Our approach is constructive.
Rather than asking an unconstrained neural network theory to discover topology dynamically, we explicitly enlarge the space of NN-FT latent variables so that continuous neural variables and discrete sector labels coexist in a single ensemble.
Schematically, for network parameters $\Theta=(\theta,Q)$ with $\theta$ continuous and $Q$ a discrete topological label, then observables take the mixed form
\be
\langle \mathcal{O} \rangle = \sum_Q \int d\theta\; P(\Theta)\,\mathcal{O}[\phi_\theta] ~,
\ee
where $P(\Theta)$ denotes the probability distribution over discrete and continuous labels.
The point of this extension is not merely formal: it provides a unified parameter space language in which smooth fluctuations and topological data can be sampled together.

We consider two case studies.
The Berezinskii--Kosterlitz--Thouless (BKT) transition is the paradigmatic finite temperature transition in two-dimensional systems with a compact $U(1)$ degree of freedom, such as the classical $XY$ model.
Its essential feature is that the low-temperature phase does not exhibit true long range order, but instead has quasi-long range order, with correlation functions that decay algebraically rather than approaching a constant.
The transition is topological: at low temperature, the relevant defects are tightly bound vortex--antivortex pairs, while at a critical temperature these pairs unbind and proliferate, producing a plasma of free vortices that screens the logarithmic interaction and drives the system into a phase with exponentially decaying correlations.
Berezinskii's original work~\cite{Berezinskii1971} established that two-dimensional systems with continuous symmetry can evade conventional ordering while still supporting a distinct low-temperature phase, and Kosterlitz and Thouless~\cite{KosterlitzThouless1973} then identified the vortex unbinding mechanism and its renormalization group description~\cite{Kosterlitz1974,Jose1977}, thereby providing the modern picture of the transition.
This is a compact boson model because the fundamental field is an angular variable, $\theta \sim \theta+2\pi$, so the target space is a circle $S^1$ instead of the real line.
This compact identification permits configurations with non-trivial winding and vortex defects, and these topological sectors are precisely what underlie the BKT transition.

In the non-linear sigma model on the worldsheet of a bosonic string propagating in background fields $(G_{\mu\nu},B_{\mu\nu},\Phi)$ when one target space direction is $S^1$, the worldsheet field corresponding to this spacetime direction is a compact boson.
Maps from the string worldsheet to the circle decompose into sectors labeled by momentum and winding.
T-duality is the statement that, for a background with an Abelian isometry along this circle direction, the sigma model admits a  dual description in which momentum and winding are exchanged and the background fields are transformed according to the Buscher rules~\cite{Buscher:1987sk,Buscher:1987qj}.
For a circle this reduces to the familiar symmetry $R\leftrightarrow \alpha'/R$, so the compact boson at radius $R$ defines the same physics as the compact boson at $\widetilde{R} = \alpha'/R$.
In more complicated settings, T-duality is the physical mechanism underlying mirror symmetry~\cite{StromingerYauZaslow1996,HoriVafa2000}.

In the BKT problem, the basic decomposition is into spin-wave fluctuations and vortex sectors.
In the T-duality problem, the analogous decomposition is into oscillator modes and momentum--winding sectors.
Both examples test the same structural claim about NN-FT: Gaussian neural degrees of freedom capture the local sector, while the full compact theory requires explicit discrete data that track its topological content.
The two case studies are complementary rather than redundant.
BKT probes the dynamical side of the question: once vortices are included, can the NN-FT ensemble access the change in phase structure associated with vortex unbinding and the departure from the Gaussian fixed line?
T-duality probes the kinematical and symmetry side: once momentum and winding are included, can the same framework realize the exact equivalence between dual compactifications and reproduce the standard exchange of topological quantum numbers and background data?
Taken together, these tests examine both the dynamics and the exact consistency conditions of compact theories in parameter space.

The organization of the paper is as follows.
In Section~\ref{sec:NN-FT}, we review NN-FT and present the mixed continuous/discrete generalization appropriate to compact targets.
In Section~\ref{sec:bkt}, we apply this philosophy to realize both sides of the Berezinskii--Kosterlitz--Thouless phase transition.
In particular, we show that a wrapped Gaussian construction reproduces the spin-wave line of the low-temperature BKT regime, while an explicit vortex sector is needed to access the disordered phase, and we reproduce the expected power law and exponential scalings of the correlation functions.
In Section~\ref{sec:td}, we show the invariance of the string theory sigma model under T-duality.
Combining Gaussian oscillator modes with explicit momentum--winding parameters and fields yields the expected compact boson T-duality identities and their extension to constant toroidal backgrounds.
We observe symmetry enhancement at the self-dual radius and apply the construction to a toy model of a T-fold.
Section~\ref{sec:disc} provides a prospectus for future directions.

\section{Neural network field theory}\label{sec:NN-FT}
The \textbf{neural network field theory} (NN-FT) program proposes that certain Euclidean field theories can be defined by specifying two pieces of data:
\begin{enumerate}
\item[(i)] a neural network architecture (a parameterized family of functions) and
\item[(ii)] a probability density on its parameters.
\end{enumerate}
Correlation functions are then computed as expectations over parameter space, rather than as functional integrals over field configurations. This idea was conceived in~\cite{Halverson:2020trp,Halverson:2021aot} and developed further in, \textit{e.g.},~\cite{Maiti:2021fpy,Demirtas:2023fir,Halverson:2024axc,Capuozzo:2025ozt,Robinson:2025ybg,Ferko:2026axm}. 
The aim of this section is to give a self-contained overview for readers with a quantum field theory background and minimal machine learning (ML) expertise.

\paragraph{From networks to random fields:}
Fix Euclidean spacetime $\mathbb{R}^d$ with coordinate $x$.
A (scalar) neural network is simply a real valued function
\be
\phi_\theta:\mathbb{R}^d\to \mathbb{R} ~.
\ee
The detailed structure of the parameterized function $\phi_\theta$ is known as the architecture. For instance, a fully-connected network a.k.a.\ a multi-layer perceptron is
built from a composition of affine maps and nonlinearities (``activations''), with trainable parameters $\theta$ (weights and biases).
We will eventually think of the initialized neural network as a map that returns the value of a field at the corresponding point in spacetime (or space, since we work in Euclidean signature).
In the NN-FT setup, the parameters $\theta$ are treated as random variables with a chosen density $P(\theta)$.
Given an ``observable'' $\mathcal{O}$ that depends on the network output, we define\footnote{In practice one often works with a product measure on parameters, \textit{e.g.}, independent and identically distributed (i.i.d.)\ random variables for weights and uniform phases, but NN-FT does not require this.}
\be\label{eq:param_expectation}
\langle \mathcal{O} \rangle_{P} = \int d\theta\; P(\theta)\,\mathcal{O}[\phi_\theta] ~.
\ee
For later applications it is useful to allow the latent data to contain both continuous and discrete pieces. If $\Theta=(\theta,Q)$, where $\theta$ is continuous and $Q$ labels a countable set of sectors (for example vortex sectors, momentum--winding data, or other topological charges), then the natural generalization is
\be
\langle \mathcal O \rangle
=
\sum_{Q}\int d\theta\; P(\theta,Q)\,\mathcal O[\phi_{\theta,Q}] ~.
\label{eq:param_expectation_mixed}
\ee
Nothing in the NN-FT formalism requires the measure to factorize, and several of the examples below will use such mixed continuous/discrete ensembles when the target theory contains explicit topological sectors that are not generated by a single-valued Gaussian covering field\footnote{A compact field $\phi ( x )$ defined on $S^1\cong \mathbb{R}/(2\pi r\mathbb{Z})$ is the physical field.
We sometimes work with a \emph{Gaussian covering field}, by which we mean a real-valued Gaussian lift to the universal cover of the compact target, namely, the noncompact field on $\mathbb{R}$ before quotienting by the circle identification.}.
In particular, the $n$-point functions are
\be\label{eq:npt_def}
G_n(x_1,\dots,x_n) = \langle \phi(x_1)\cdots \phi(x_n)\rangle_{P} ~,
\ee
where we suppress the $\theta$-dependence inside the expectation.

This is reminiscent of a path integral, but the integration is over parameter space rather than over field configurations. 
The output $\phi(x)$ is a random field on $\mathbb{R}^d$ whose statistics are induced from $P(\theta)$ and the architecture. 
If the resulting Schwinger functions $G_n$ satisfy~\cite{Halverson:2021aot} the Osterwalder--Schrader axioms~\cite{Osterwalder:1973dx,Osterwalder:1974tc}, then they define a bona fide Euclidean quantum field theory (and can be analytically continued to a Lorentzian theory with unitarity replacing reflection positivity).
In one dimension, remarkably broad existence results are known: under mild assumptions, essentially any Osterwalder--Schrader framework consistent quantum mechanics admits a neural network representation~\cite{Ferko:2025ogz}.

\paragraph{Infinite width and the emergence of free fields:}
A central technical input is that many architectures have large-$N$ limits in which their outputs become Gaussian processes (GPs), \textit{i.e.}, free fields in quantum field theory language.
For concreteness, consider a single hidden layer ``feature expansion''\footnote{This is a standard starting point in machine learning and is also the natural arena for the classic neural network / Gaussian process correspondence~\cite{neal,williams}.}
\be\label{eq:one_layer}
\phi_N(x) = \frac{1}{\sqrt{N}} \sum_{i=1}^{N} a_i\,\phi(x;w_i) ~,
\ee
where $N$ is the width (number of hidden units), $\phi(x;w)$ is a feature map (\textit{e.g.}, $\phi=\sigma(w\cdot x + b)$ for an activation $\sigma$), and the parameters $\{a_i,w_i\}$ are i.i.d.\ draws from a fixed distribution.
For fixed $x$, $\phi_N(x)$ is a sum of $N$ independent random variables.
Under mild moment conditions, the multivariate central limit theorem implies that for any finite set of points $\{x_\alpha\}$ the vector $\{\phi_N(x_\alpha)\}$ becomes jointly Gaussian as $N\to\infty$.
Equivalently, in the infinite width limit the random field $\phi(x)$ is a Gaussian process with two-point function
\be\label{eq:kernel_def}
K(x,x'):= \lim_{N\to\infty}\langle \phi_N(x)\phi_N(x')\rangle = \mathbb{E}_{w,a}\left[a^2\,\phi(x;w)\phi(x';w)\right] ~,
\ee
and vanishing connected correlators of order $n>2$.
This is the neural network Gaussian process (NNGP) correspondence~\cite{neal,williams,Matthews2018GaussianPB}. 
Deep networks (multiple layers) and other architectures also admit Gaussian process limits, though the induced kernel $K$ can become more structured~\cite{schoenholz2017correspondence,yangTP1}.  

From the quantum field theory viewpoint,~\eref{eq:kernel_def} is nothing but a free theory with covariance $K$:
\be\label{eq:gaussian_measure}
\langle \cdots \rangle_\text{GP} \propto \int [\mathcal{D}\phi]\; \exp\left[-\frac12\int d^dx\,d^dy\, \phi(x)\,K^{-1}(x,y)\,\phi(y)\right] (\cdots) ~.
\ee
An NN-FT idea is to engineer $K$  by choosing the architecture and $P(\theta)$ so that the induced Schwinger functions match those of a target field theory~\cite{Halverson:2021aot,Demirtas:2023fir}.

\paragraph{Engineering propagators via random feature architectures:}
The simplest bridge to quantum field theory is obtained by choosing features that are Fourier modes.
Take
\be\label{eq:rff_arch}
\phi_N(x) = \frac{1}{\sqrt{N}}\sum_{i=1}^N a_i \cos(w_i\cdot x + b_i) ~,
\quad a_i\sim \mathcal{N}(0,\sigma_a^2),\quad  b_i\sim \mathrm{Unif}[0,2\pi] ~,
\ee
with momenta $w_i$ drawn from a density $\rho(w)$ on $\mathbb{R}^d$.
Averaging over $a_i$ and $b_i$ gives
\be\label{eq:rff_kernel}
K(x,x') = \lim_{N\to\infty}\langle \phi_N(x)\phi_N(x')\rangle = \frac{\sigma_a^2}{2}\int_{\mathbb{R}^d} d^dw\;\rho(w)\,\cos\big(w\cdot(x-x')\big) ~.
\ee
This is the random feature construction familiar from kernel methods~\cite{rahimi2007random}, but here it is repurposed as a definition of a free field~\cite{Halverson:2021aot}. 
If $\rho(w)$ is rotationally invariant, then $K(x,x')$ depends only on $|x-x'|$, and translation/rotation invariance of the induced ``field theory'' is manifest.

We observe that~\eref{eq:rff_kernel} makes the match to standard propagators transparent in momentum space.
Up to normalization conventions, the Fourier transform of $K$ is proportional to the spectral density $\rho$.
Therefore, by choosing
\be\label{eq:spectral_choice}
\rho(w) \propto \frac{1}{w^2+m^2}\;\;\times\;\; \bm{1}_{\epsilon<|w|<\Lambda} ~,
\ee
one obtains a covariance $K(p)\propto (p^2+m^2)^{-1}$ for momenta between an infrared (IR) cutoff $\epsilon$ and ultraviolet (UV) cutoff $\Lambda$.
In this sense, the propagator of a free scalar field can be realized as an ensemble average over network parameters.
More elaborate choices of features and parameter distributions can realize fermionic propagators and additional internal structure~\cite{Huang:2025ipy,Frank:2025zuk}.

\paragraph{Interactions as finite width effects and non-Gaussianities:}
Real networks have finite width, which introduces interactions as non-Gaussianities.
At large but finite $N$, connected correlators scale with powers of $1/N$ and can be organized systematically~\cite{Yaida2019NonGaussianPA,PhysRevE.104.064301,roberts2022principles,Sen:2025vzl}.
From a quantum field theory perspective, this is analogous to a large-$N$ expansion: $N\to\infty$ gives a free theory, while $1/N$ corrections generate interaction vertices.

There are two complementary ways interactions enter:
\begin{itemize}
\item \textbf{Finite width corrections at fixed parameter prior:}  
Even when $P(\theta)$ is a simple product of Gaussians, finite width induces nonzero higher cumulants of $\phi(x)$, hence non-trivial connected $n$-point functions.  
These can be interpreted as arising from effective interaction terms in an induced action for $\phi$~\cite{Demirtas:2023fir}.
\item \textbf{Non-Gaussian parameter densities:}
One may choose a non-Gaussian or non-factorized $P(\theta)$ directly, thereby building interactions into parameter space.
A key question is when the resulting effective action for $\phi$ is local (or approximately local) in spacetime.
Progress on actions and locality criteria was developed in~\cite{Demirtas:2023fir}.
\end{itemize}
Concrete NN-FT realizations of $\phi^4$ theory, Yukawa couplings, and supersymmetric interacting theories have been constructed along these lines~\cite{Demirtas:2023fir,Frank:2025zuk}.

\paragraph{Symmetries and targets:}
Symmetries are implemented by design: one builds the desired invariances into the architecture and/or into $P(\theta)$.
A useful organizing principle is symmetry via duality, wherein one designs parameter space densities that reproduce invariant correlators~\cite{Maiti:2021fpy}.
This has been applied to construct Euclidean invariant NN-FTs~\cite{Halverson:2021aot} and to more structured symmetry targets.
Recent developments include conformal field theories from neural networks~\cite{Halverson:2024axc}, conformal defects~\cite{Capuozzo:2025ozt},  realizations of full Virasoro symmetry in two dimensions~\cite{Robinson:2025ybg}, and the computation of the Veneziano and Virasoro--Shapiro amplitudes for the bosonic string~\cite{Frank:2026bui}.

Although NN-FT is a physics motivated program, it draws on broad machine learning theory about infinite size limits.
For example, convolutional networks and attention mechanisms~\cite{vaswani2017attention} also admit NNGP/neural tangent kernel descriptions in appropriate limits~\cite{Novak2018BayesianCN,GarrigaAlonso2019DeepCN,hron2020infinite,yangTP2,Jacot2018NeuralTK}.  
For a physics oriented introduction to these ideas, see the lecture notes~\cite{Halverson:2024hax}.

\paragraph{Summary:}
NN-FT replaces the usual ``sum over fields'' with a ``sum over network parameters,'' in which the architecture and parameter density play roles analogous to kinematics and dynamics.
Infinite width limits reproduce free fields through the NNGP correspondence, while interactions arise from controlled non-Gaussianities.  
This provides a new and technically tractable route to defining and computing correlators in quantum field theory, and it opens a channel for importing tools from modern machine learning theory into field theoretic model building.

\section{Berezinskii--Kosterlitz--Thouless transition}\label{sec:bkt}

The Berezinskii--Kosterlitz--Thouless (BKT) transition is the paradigmatic two-dimensional phase transition driven by the proliferation of topological defects rather than by conventional symmetry  breaking~\cite{Berezinskii1971,KosterlitzThouless1973,Kosterlitz1974}.
Its canonical realization is the two-dimensional XY model,
\be
H_\text{XY}=-J\sum_{\langle ij\rangle}\cos(\theta_i-\theta_j) ~,
\ee
with a compact angle variable $\theta\sim \theta+2\pi$.
Passing from the lattice to the continuum replaces the variable $\theta_i$ by a compact scalar field $\theta(x)\sim \theta(x)+2\pi$. Its infrared action is
\be
S_\text{sw}=\frac{K_R}{2}\int d^2x\,(\nabla\theta)^2 ~,
\qquad K_R :=\frac{\rho_{s,R}}{T} ~,
\label{eq:spinwave}
\ee
which is Gaussian, and where $\rho_{s,R}$ is the renormalized spin stiffness (superfluid stiffness in the superfluid/superconductor interpretation).
$K_R$ is the dimensionless renormalized stiffness.
As usual, the renormalized quantities are obtained after integrating out short distance physics.
We should regard~\eref{eq:spinwave} as an effective action of a continuum theory that gives the same physics as the BKT transition in the XY model.

Because continuous symmetries cannot be spontaneously broken in two dimensions at finite temperature~\cite{MerminWagner1966}, the ordered phase is replaced by a critical phase with quasi-long range order and algebraic correlations.
To see this, we consider the vertex operators
\be
V_q(x) = e^{iq\theta(x)} ~, \qquad q\in\mathbb{Z} ~.
\label{eq:vertex_operator_def}
\ee
The physically relevant correlator is
\be
G_q(x):=\langle V_q(x)V_{-q}(0)\rangle = \Big\langle e^{iq(\theta(x)-\theta(0))}\Big\rangle ~.
\label{eq:full_compact_correlator}
\ee

In the low-temperature phase, vortices are bound into neutral pairs and only renormalize the stiffness.
The infrared theory lies on the Gaussian fixed line.
Accordingly,
\be
G_q(x)\sim |x|^{-q^2/2\pi K_R(T)} ~, \qquad T<T_c ~.
\label{eq:algebraic_compact_correlator}
\ee
The theory is gapless in the infrared and exhibits quasi-long range order rather than true spontaneous symmetry breaking.
In high-energy language, this phase is a line of $c=1$ compact boson fixed points with continuously varying scaling dimensions.
Vortices are point topological defects around which $\theta$ winds by $2\pi m$ ($m\in\mathbb{Z}$).
An isolated vortex of charge $m$ has an energy cost that grows logarithmically with system size, $L$:
\be
\frac{E_m}{T} \sim \pi K_R\, m^2\, \log(L/a) ~. \label{eq:bf}
\ee
Here, $a$ is the short distance cutoff, generally identified with the vortex core radius, at which the continuum description ceases to be valid.
The conclusion we draw from~\eref{eq:bf} is that free vortices are suppressed at low-$T$~\cite{KosterlitzThouless1973,Kosterlitz1974}.

It is useful to decompose the field into a smooth part and a singular vortex part,
\be
\theta(x)=\theta_\text{sw}(x)+\theta_v(x) ~.
\label{eq:theta_sw_vortex_split}
\ee
This decomposition is conceptually helpful, but only the full compact operator is the physical observable.
The smooth piece by itself remains a massless Gaussian field, so the two-point function of $e^{iq\theta_\text{sw}(x)}$ follows power law scaling as in~\eref{eq:algebraic_compact_correlator}.
What changes across the BKT transition is not the existence of the Gaussian sector, but the infrared behavior of the full compact correlator once vortex configurations are included.

At sufficiently high temperature, entropy wins over the logarithmic energy, and vortices unbind.
The system is then a Coulomb gas of free topological charges, which screens correlations.
The full two-point function then becomes short ranged with a finite correlation length $\xi(T)$.

The continuum way to see this is through the dual sine--Gordon description,
\be
S_\text{dual} = \frac{1}{2\pi K}\int d^2x\,(\nabla\widetilde\theta)^2 -2y\int d^2x\,\cos\widetilde\theta ~,
\label{eq:dual_sine_gordon}
\ee
where $y$ is the vortex fugacity.\footnote{\samepage
A convenient dualization is obtained by introducing an auxiliary current
$j^\mu$:
$$
\exp\left[-\frac{K}{2}\int d^2x\ (\partial_\mu\theta)^2 \right] \propto
\int [\mathcal{D}j^\mu]\, \exp\left[-\frac{1}{2K}\int d^2x\ j_\mu j^\mu +i\int d^2x\ j^\mu\,\partial_\mu\theta \right] ~.
\label{eq:HS_current_dual}
$$
Here and in~\eref{eq:dual_sine_gordon}, $K$ denotes the scale-dependent stiffness in the dual description; in the low-temperature phase its infrared limit is $K_R$.
Integrating over the smooth field $\theta_\text{sw}$ imposes current
conservation, $\partial_\mu j^\mu=0$, which in two Euclidean dimensions is solved locally by $j^\mu=\frac{1}{2\pi}\,\epsilon^{\mu\nu}\partial_\nu\varphi$.
Substituting this into the above gives 
$$
S_\text{dual}[\varphi;\{q_a\}] = \frac{1}{2\pi^2 K}\int d^2x\,(\nabla\varphi)^2
-i\sum_a q_a\,\varphi(x_a) ~,
\label{eq:dual_action_sources}
$$
up to an overall normalization.
The last term shows that a unit vortex insertion is created by $e^{\pm i\varphi}$.
Summing over a dilute neutral gas of unit vortices, $q_a=\pm1$, exponentiates
these insertions.
After the field redefinition $\varphi=2\widetilde\theta$, one obtains the standard sine--Gordon form.
}
The precise normalization of $\widetilde\theta$ is conventional, but the cosine term represents unit vortex insertions and becomes marginal precisely at the BKT transition~\cite{Kosterlitz1974,Jose1977}.
Below $T_c$, this perturbation is irrelevant and the theory flows back to the gapless compact boson.
Above $T_c$, the vortex fugacity is relevant, the cosine flows to strong coupling, and a mass scale is dynamically generated.
The physical electric correlator~\eref{eq:full_compact_correlator} then crosses over to
\be
G_q(x)\sim e^{-|x|/\xi(T)} ~, \qquad T>T_c ~,
\label{eq:massive_compact_correlator}
\ee
with $\xi(T)$ diverging as $T\to T_c^+$.
Equivalently, the magnetic disorder operators proliferate and screen the electric vertex operators.

This is the sense in which the BKT transition is topological rather than Landau-like.
There is no local order parameter that acquires an expectation value on one side and vanishes on the other.
Instead, the infrared behavior of correlation functions changes qualitatively:
below $T_c$, the two-point function is algebraic, while above $T_c$ it is exponentially decaying.
This is the key continuum observable our NN-FT construction should reproduce.

The BKT transition is often called ``infinite order'' because the singularity in the free energy is essential rather than algebraic: the correlation length diverges as
\be
\xi(T)\sim a\,\exp\left(\frac{c}{\sqrt{(T-T_c)/T_c}}\right) ~, \qquad T\to T_c^+ ~,
\label{eq:xi_essential}
\ee
so the free energy is non-analytic but all derivatives at $T_c$ can remain finite (no latent heat, no finite order power law singularity)~\cite{Kosterlitz1974,Minnhagen1987}.
Physically, this reflects the fact that the transition is controlled by the (un)binding of vortices: a discrete change in the defect plasma, not in a local order parameter.

The running couplings $K(\ell)$ and $y(\ell)$ obey the BKT flow equations.
In terms of $\ell=\log\mu$, the renormalization group ``time,'' we have the evolution equations~\cite{Kosterlitz1974,Jose1977}
\be
\frac{dK^{-1}}{d\ell}=4\pi^3y^2+\cdots ~,
\qquad \frac{dy}{d\ell}=\bigl(2-\pi K\bigr)y+\cdots ~.
\label{eq:KT_RG}
\ee
The renormalized stiffness is the infrared limit $K_R=\lim_{\ell\to\infty}K(\ell)$ when that limit exists.
If the system starts with $K > 2/\pi$ and small $y$, the flow drives $y\to 0$.
The vortices are rendered irrelevant, and the system flows to a line of fixed points corresponding to a stable superfluid phase with a finite macroscopic stiffness $\Upsilon$.
If the system starts with $K < 2/\pi$, the flow drives $y\to\infty$.
Vortices proliferate, destroying the stiffness and driving $K\to 0$ (the disordered phase).
The line of Gaussian fixed points at $y=0$ (the spin-wave CFT) is parameterized by $K$.
The BKT transition occurs when the vortex perturbation is marginal, $\pi K_c=2$, so that
\be
\eta(T_c)=\frac{1}{2\pi K_R(T_c)}=\frac14 ~.
\label{eq:eta_c}
\ee
and
\be
\rho_{s,R}(T_c^-)=\frac{2T_c}{\pi} ~.
\label{eq:universal_jump}
\ee

In addition to the two-point function having different behavior on the two sides of the transition, the \emph{helicity modulus} is a macroscopic quantity that physically distinguishes the low-temperature and high-temperature phases.
For the XY model, the helicity modulus is defined by twisting the boundary condition, or equivalently by coupling the phase field to a uniform background gauge field $A_x=\Phi/L$.
One may write
\be
\Upsilon_\text{twist} =
\frac{1}{L^2} \left. \frac{\partial^2 F(\Phi)}{\partial (\Phi/L)^2} \right|_{\Phi=0} ~.
\ee
On the lattice, this can be written exactly as a diamagnetic minus paramagnetic response,
\be
\Upsilon_\text{twist} =
\frac{J}{L^2} \Bigl\langle \sum_{\mathbf x}\cos(\Delta_x\theta)\Bigr\rangle -
\frac{J^2}{T L^2} \Biggl\langle \Bigl(\sum_{\mathbf x}\sin(\Delta_x\theta)\Bigr)^2 \Biggr\rangle ~.
\ee
This is an exact observable of the finite system and does not assume a Gaussian effective action.
Historically, this theoretical prediction provided the most rigorous experimental test for the BKT transition and was famously confirmed in thin films of liquid Helium-4 by Bishop and Reppy~\cite{bishop1978study}.

In the continuum description, where we rely on effective field theory, the long-distance physics is described by
\be
F_\text{eff}[\theta;A] =
\frac{\Upsilon_R}{2}\int d^2x\,(\nabla\theta-A)^2 ~,
\ee
and dividing by temperature gives the dimensionless action
\be
S_\text{eff}[\theta;A] =
\frac{1}{T}F_\text{eff} =
\frac{K_R}{2}\int d^2x\,(\nabla\theta-A)^2 ~,
\qquad K_R=\frac{\Upsilon_R}{T} ~.
\ee
In that same Gaussian infrared theory, the phase Green's function obeys
\be
\langle |\theta_k|^2\rangle \sim \frac{1}{k^2\, K_R} ~,
\ee
so that
\be
\Upsilon_R = T K_R ~. \label{eq:ups}
\ee
Indeed, the renormalized quantity $\Upsilon_R(T_c^-)$ is the expression~\eref{eq:universal_jump}.
The helicity modulus and the Green's function stiffness are the same coefficient in the infrared effective theory.

The free vortices act analogously to free charges in a dielectric medium.
When a global phase twist is applied, the free vortices move perpendicularly to the current, causing phase slips.
This perfectly screens the applied twist, driving the macroscopic resistance to zero.
Therefore, in the high-temperature phase, we expect $\Upsilon_R \to 0$.

We are careful to spell out the distinction between $\Upsilon_\text{twist}$ and $\Upsilon_R$ because the former is exact at finite size, whereas the spectral estimate assumes that the low-momentum sector is already well described by a Gaussian theory.
Below $T_c$, on sufficiently large systems, this identification is well motivated and tracks the true helicity modulus closely.
However, it is still not the same estimator.
Finite size effects, the use of only a few low-lying modes, non-Gaussian corrections, and compactness/vortex effects can all lead to deviations between the true finite size twist response and the quantity inferred from the spectrum.
Above $T_c$, this becomes even more important: in the thermodynamic limit the true helicity modulus vanishes, while a finite box may still exhibit a nonzero low-$k$ effective stiffness.

\subsection{NN-FT architecture}\label{sec:bkt_arch}

We realize the compact boson on an $L\times L$ torus via two independent sectors: a Gaussian spin-wave network and an explicit Coulomb gas vortex sampler.
Let us explain.

\paragraph{Spin-wave sector and random Fourier features:}
The covering field\footnote{
In the infinite width limit of the NN-FT, such a field is naturally produced by a scale invariant log-kernel (random Fourier feature) architecture.
As we shall see, the Gaussian covering field captures the smooth local, spin-wave-type fluctuations, but by itself it does not generate the full compact topological content of the theory. 
This is because the globally single-valued lift $\phi$ has no non-trivial winding/vortex sectors on its own; these have to be added separately or represented by patching/multivalued data.
}
$\theta_\text{sw}(x)\in\mathbb{R}$ is drawn from a random Fourier feature (RFF) network with $N$ neurons,
\be
\theta_\text{sw}(x) = \frac{A}{\sqrt{N}}\sum_{j=1}^{N}\cos(k_j\cdot x + \gamma_j) ~,
\label{eq:rff_theta_sw}
\ee
where $A=\sqrt{Z/\pi}$ is a normalization constant (see below).
The random parameters are:
\begin{itemize}\itemsep0pt
\item \textbf{Wavevector:}
  $k_j = \frac{2\pi}{L}\,n_j$, with integer mode
  $n_j=(n_x,n_y)$ drawn from
  $p(n) \propto 1/|n|^2$
  in the shell $n_{\min}^2 \le |n|^2 \le n_{\max}^2$.
\item \textbf{Phase:} $\gamma_j\sim\text{Uniform}[0,2\pi]$.
\end{itemize}
All $(n_j,\gamma_j)$ are independent across neurons.
The normalization is
$Z=\sum_{n_{\min}^2\le|n|^2\le n_{\max}^2} 1/|n|^2$.
Each draw of the $N$ neuron parameters defines one field configuration;
ensemble averages over networks may be used to compute correlators.

The XY spin field is the compact projection to $S^1$
\be
\bm{s}(x) = \big(\cos(b\,\theta_\text{sw}(x)),\; \sin(b\,\theta_\text{sw}(x))\big) ~,
\label{eq:s_def}
\ee
where $b>0$ is a tunable coupling that we will see is related to the temperature $T$.
Casting this into CFT language, since $\bm{s}(x)\cdot\bm{s}(0) = \cos\big(b(\theta_\text{sw}(x)-\theta_\text{sw}(0))\big) = \text{Re}\,e^{ib(\theta_\text{sw}(x)-\theta_\text{sw}(0))}$, we work directly with vertex operators $V_b(x) = e^{ib\,\theta_\text{sw}(x)}$ henceforth.
The subscript $b$ denotes the radius of the circle.

\paragraph{Vortex sector --- the Coulomb gas:}
The RFF field $\theta_\text{sw}(x)$ is a smooth, single-valued Gaussian --- it cannot support topological winding.
To access the high-temperature phase we augment the field with an explicit vortex sector:
\be
\theta(x) = b\,\theta_\text{sw}(x) + \theta_v(x) ~,
\label{eq:theta_decomp}
\ee
where $\theta_\text{sw}$ is the RFF spin-wave field and
\be
\theta_v(x) = \sum_{a=1}^{N_v} m_a \,\arg(x - x_a)
\label{eq:theta_vortex}
\ee
is the singular vortex field generated by a neutral gas of unit charges $m_a=\pm1$ at positions $x_a\in[0,L)^2$.
The overall neutrality constraint $\sum_a m_a = 0$ is enforced at each step. The vortex configurations are sampled from the grand canonical Coulomb gas distribution
\be
P_\text{vort}(\mathcal V) \propto y^{N_v} \exp\!\left[2\pi K_0 \sum_{a<b} m_a m_b \log(r_{ab}+a_c)\right] ~,
\label{eq:coulomb_gas}
\ee
where $y$ is the fugacity, $a_c$ is the core radius and $r_{ab}$ is the periodic distance on the torus.
The Coulomb gas coupling is tied to the bare input stiffness $K_0$ fixed by $b$.
Sampling is by the grand canonical Metropolis--Hastings algorithm.
At each step one of three moves is proposed:
(i) displacing a single vortex by a Gaussian step,
(ii) inserting a neutral $(+1,-1)$ pair at random positions (with a bias toward local placement to improve acceptance at large~$K$), or
(iii) deleting a randomly chosen $(+1,-1)$ pair.
Each proposal is accepted or rejected with the standard Metropolis ratio so as to satisfy detailed balance with respect to~\eref{eq:coulomb_gas}.
The number of vortices $N_v$ therefore fluctuates, and the fugacity $y$ controls their mean abundance: at small $b$ (large $K_0$) the logarithmic repulsion suppresses pair creation, while at large $b$ (small $K_0$) vortex--antivortex pairs proliferate, as seen in numerics.

A single $\theta$ field draw consists of one RFF configuration $\theta_\text{sw}$ together with one vortex configuration $\mathcal{V}$.
Since the two sectors are independent, the expectation of any observable
$\mathcal{O}[\theta]$ factorizes as
\be
\langle\mathcal{O}\rangle_\text{BKT}
= \int d\Theta_\text{sw}\,d\mathcal{V}\;
  P_\text{sw}(\Theta_\text{sw})\,
  P_\text{vort}(\mathcal{V})\;
  \mathcal{O}\big[\theta_{\Theta_\text{sw},\mathcal{V}}\big] ~.
\label{eq:mixed_ensemble}
\ee

\subsection{Analytic matching}\label{sec:bkt_matching}

The free correlator in the spin-wave sector can be computed exactly at finite-$N$ and finite~$L$.
By expanding the product of independent neurons and averaging over the random phases $\gamma_j$ and modes $n_j$, one obtains
\be
\langle\theta_\text{sw}(x)\,\theta_\text{sw}(0)\rangle
= \frac{1}{2\pi}\sum_{n_{\min}^2\le|n|^2\le n_{\max}^2}
  \frac{\cos\!\big(\frac{2\pi\,n\cdot x}{L}\big)}{|n|^2} ~.
\label{eq:theta_sw_2pt_exact}
\ee
This is the exact two-point function on the torus at finite cutoffs;
the field is continuous, and the discrete sum reflects the periodicity of the torus rather than any lattice.
In the large volume limit, the sum over integer modes becomes an integral in polar coordinates,
\be
\sum_n\frac{\cos(2\pi n\cdot x/L)}{|n|^2}
\to
\int n\,dn\,d\alpha\;\frac{\cos(2\pi n\,r\cos\alpha/L)}{n^2}
= 2\pi\int \frac{dn}{n}\;J_0\!\Big(\frac{2\pi n\,r}{L}\Big) ~,
\label{eq:continuum_limit}
\ee
where the angular integral produces $2\pi J_0$ from the cosine; see~\cite{Frank:2026bui} for the direct continuum calculation.
Evaluating the radial integral in the regime $a \ll r \ll L$ gives
\be
\langle\theta_\text{sw}(x)\,\theta_\text{sw}(0)\rangle
\xrightarrow{L,\,n_{\max}\to\infty}
-\log|x| + \dots
\label{eq:theta_sw_propagator}
\ee
matching the expected $2d$ log correlator.

The spin-spin correlator is computed as follows.
From~\eref{eq:rff_theta_sw},
$\Delta\theta_\text{sw}:=\theta_\text{sw}(x)-\theta_\text{sw}(0) = \frac{A}{\sqrt{N}}\sum_{j=1}^{N}\delta_j$,
where
$\delta_j = \cos(k_j\cdot x + \gamma_j)-\cos(\gamma_j)$
are i.i.d.\ random variables.
The characteristic function therefore factors over neurons,
\be
\big\langle e^{ib\Delta\theta_\text{sw}}\big\rangle
= \prod_{j=1}^{N}\Big\langle e^{i\frac{bA}{\sqrt{N}}\delta_j}\Big\rangle
= \Big[\big\langle e^{it\delta}\big\rangle\Big]^N ~,
\qquad t \equiv \frac{bA}{\sqrt{N}} ~.
\label{eq:char_fn_factor}
\ee
Expanding for large-$N$ ($t\to 0$):
$\langle e^{it\delta}\rangle = 1 - \frac{t^2}{2}\langle\delta^2\rangle + O(t^3)$,
with $\langle\delta\rangle=0$ by the phase average.
Using $(1+c/N)^N\to e^c$ gives
\be
\big\langle e^{ib\Delta\theta_\text{sw}}\big\rangle
\xrightarrow{N\to\infty}
\exp\!\Big[-\frac{b^2}{2}\big\langle(\Delta\theta_\text{sw})^2\big\rangle\Big]
= \exp\!\Big[-\frac{b^2}{2\pi}
  \sum_n\frac{1-\cos(2\pi n\cdot x/L)}{|n|^2}\Big] ~,
\label{eq:vertex_exact}
\ee
where the last equality substitutes
$\langle(\Delta\theta_\text{sw})^2\rangle
= 2[\langle\theta_\text{sw}^2\rangle - \langle\theta_\text{sw}(x)\theta_\text{sw}(0)\rangle]$
using~\eref{eq:theta_sw_2pt_exact}.
The finite torus formula~\eref{eq:vertex_exact} is exact at $N=\infty$.
Using the continuum limit result~\eref{eq:continuum_limit}--\eref{eq:theta_sw_propagator},
and noting that
$\langle(\Delta\theta_\text{sw})^2\rangle \to 2\log|x| + \text{const}$,
we obtain
\be
\big\langle V_b(x)\,V_b^*(0)\big\rangle
= \big\langle e^{ib(\theta_\text{sw}(x)-\theta_\text{sw}(0))}\big\rangle
\sim |x|^{-b^2} ~.
\label{eq:eta_result}
\ee
The dimension of the NN-FT spin field is therefore
\be
\eta_\text{sw}(b) = b^2 ~.
\label{eq:eta_b2}
\ee

Matching to the XY model via~\eref{eq:eta_c} and~\eref{eq:universal_jump}, $\eta = T/(2\pi\rho_s)$, gives the dictionary
\be
K_0 = \frac{1}{2\pi b^2} ~, \qquad
T = \frac{\rho_{s,0}}{K_0} = 2\pi\rho_{s,0}\, b^2 ~, \qquad
b_c = \frac{1}{2}
\quad(\text{from } \eta_c = \tfrac{1}{4}) ~.
\label{eq:bkt_eta}
\ee
In performing the matching in the low-temperature phase we choose units in which the bare stiffness $\rho_{s,0}=1$ because the vortex contribution has not yet renormalized the stiffness.
In the spin-wave only sector there is no vortex screening, so $K_R=K_0$; once the vortex sector is included, the infrared stiffness measured below is the renormalized quantity $K_R$, not the bare input $K_0$.
The NN-FT spin-wave only sector realizes the Gaussian critical line with the dimensionless stiffness $K_R=K_0=1/(2\pi b^2)$.
The BKT transition occurs at $b_c=1/2$, where $T_c = \pi/2$.

Because the spin-wave and vortex sectors are sampled independently, the full two-point function factorizes:
\be
G_2(r) = \big\langle\cos(\theta(x)-\theta(0))\big\rangle
= G_\text{sw}(r) \times G_v(r) ~,
\label{eq:G2_factorization}
\ee
where $G_\text{sw}\sim r^{-b^2}$ is the spin-wave contribution and $G_v(r)$ is the vortex contribution.
Below $T_c$, vortices are bound and $G_v(r)\to\text{const}$;
above $T_c$, vortex proliferation suppresses $G_v$ and drives the system to disorder.

A useful bridge to the standard lattice treatment is the Villain form of the XY model,
\be
 e^{K_0\cos(\Delta\theta)} \approx \sum_{m\in\mathbb Z}
 \exp\!\left[-\frac{K_0}{2}(\Delta\theta-2\pi m)^2\right].
\ee
Dualizing the Villain form yields a Gaussian spin-wave sector together with a neutral Coulomb gas of integer vortices.
Our mixed ensemble is designed to reproduce that same structure at the continuum level: the RFF field supplies the smooth Gaussian sector, while $P_\text{vort}$ samples the corresponding vortex gas.
In that sense the construction is Villain-inspired, and the factorization in~\eref{eq:G2_factorization} is exact within our two-sector sampling scheme.

\subsection{Numerical results}\label{sec:bkt_numerics}

We present a systematic suite of measurements on an $L=64$ torus with $N_\text{features}=256$ RFF modes and fugacity $y=0.002$.
Each run uses $N_\text{samples}=300$ measured configurations.
For observables involving vortices, the spin-wave sector is resampled independently for each measurement, while the vortex sector is generated by a thermalized Metropolis--Hastings chain with $150$ update steps between successive samples.

\begin{figure}[t]
\centering
\includegraphics[width=0.85\textwidth]{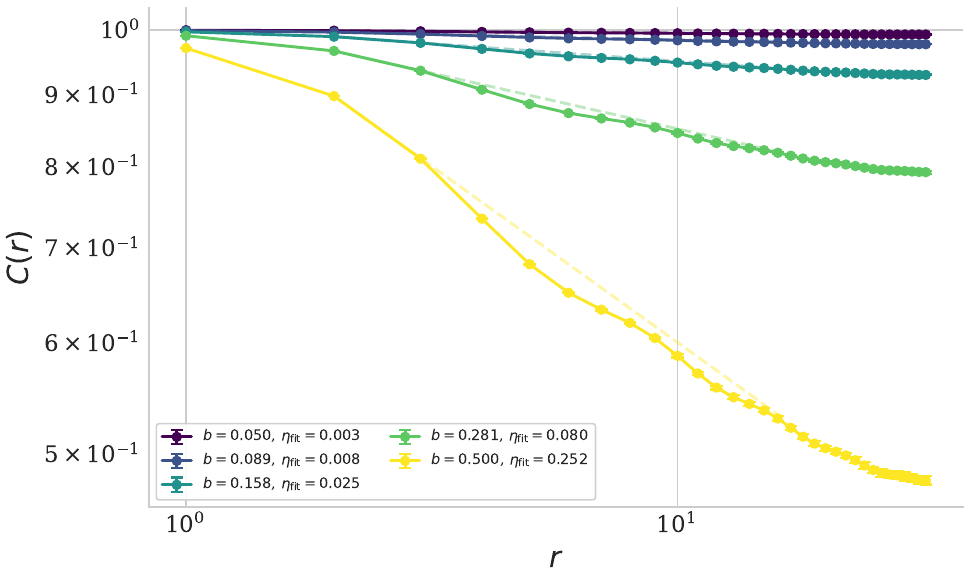}
\caption{Spin-wave correlator $C(r)$ below $T_c$ (log-log).
Solid lines: data.
Dashed lines: predicted power law $r^{-b^2}$.
\label{fig:corr_below}}
\end{figure}

\paragraph{Spin-wave correlator below $T_c$:}
Figure~\ref{fig:corr_below} shows the spin-wave correlator $C(r)=\langle\cos\big(b(\theta_\text{sw}(x)-\theta_\text{sw}(x+r))\big)\rangle$ for five values $b\le b_c$ in log-log coordinates.
At each~$b$, a power law fit $C(r)\sim r^{-\eta_\text{fit}}$ is performed in the window $r\in[2,L/4]$.

\begin{table}[t]
\centering
\begin{tabular}{cccc}
\hline
$b$ & $\eta_\text{exp}=b^2$ & $\eta_\text{fit}$ & $R^2$ \\
\hline
0.050 & 0.00250 & 0.00252 & 0.99196 \\
0.089 & 0.00791 & 0.00796 & 0.99196 \\
0.158 & 0.02500 & 0.02517 & 0.99196 \\
0.281 & 0.07906 & 0.07963 & 0.99198 \\
0.500 & 0.25000 & 0.25204 & 0.99204 \\
\hline
\end{tabular}
\caption{Below-$T_c$ power law exponents.
$\eta_\text{fit}$ agrees with $b^2$ to better than $1\%$ throughout, confirming~\eref{eq:eta_b2}.
\label{tab:below_Tc}}
\end{table}

Table~\ref{tab:below_Tc} compares the fitted exponents to the analytic prediction~$\eta=b^2$.
The agreement is excellent: for all values tested, $\eta_\text{fit}/b^2 \approx 1$ with $R^2>0.99$.
At the critical point $b=0.5$, the measured exponent $\eta_\text{fit}\approx 0.25$
matches the universal BKT value~\eref{eq:universal_jump}.

\begin{figure}[t]
\centering
\includegraphics[width=0.85\textwidth]{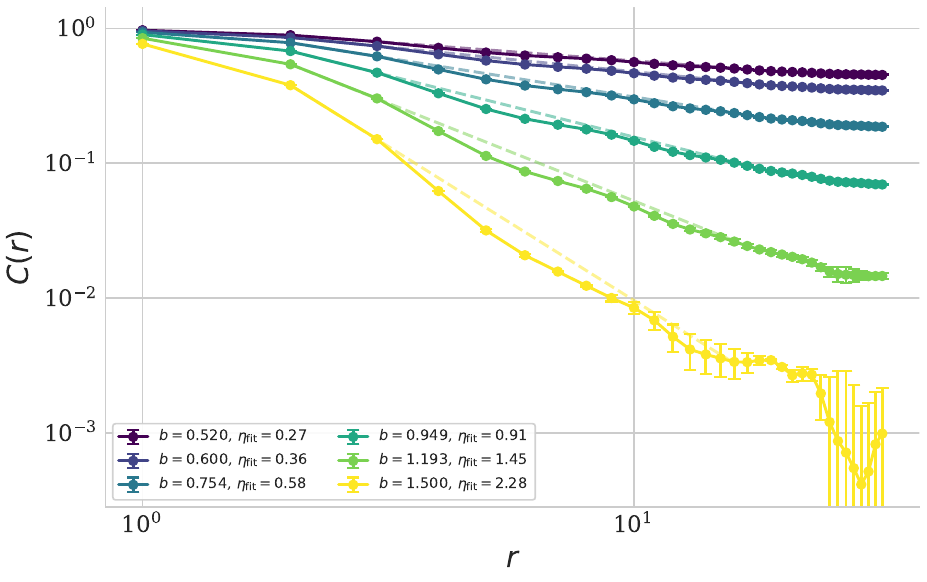}
\caption{Spin-wave correlator above $T_c$ (log-log).
The correlator remains power law because the RFF field is Gaussian by construction: the spin-wave sector alone has no disordered phase.
\label{fig:corr_above}}
\end{figure}

\paragraph{Spin-wave correlator above $T_c$:}
Figure~\ref{fig:corr_above} shows the same spin-wave correlator for $b>b_c$.
In the pure spin-wave sector, the field $\theta_\text{sw}$ is Gaussian at every~$b$: there is no mechanism for generating a mass gap.
Accordingly, $C(r)$ continues to exhibit power law decay $C(r)\sim r^{-b^2}$ for $b>b_c$, with the fitted exponents tracking the analytic prediction.

This is the expected result: the spin-wave-only theory realizes the entire Gaussian critical line, with no transition.
The onset of disorder requires the vortex sector.

\paragraph{Full correlator:}\label{sec:corr_full}

We also computed the full two-point function $G_2(r) = \langle\cos(\theta(x)-\theta(x+r))\rangle$ with $\theta = b\,\theta_\text{sw} + \theta_v$, spanning both $b<b_c$ and $b>b_c$.
Although we do not display a separate figure, the data behave as expected.
Below $T_c$, the vortex factor $G_v(r)\approx\text{const}$, so the full correlator exhibits the same power law behavior as the spin-wave sector.
Above $T_c$, vortex proliferation suppresses $G_2$ relative to $G_\text{sw}$, signaling the onset of disorder.
The factorization~\eref{eq:G2_factorization} therefore makes the role of each sector transparent: the Gaussian spin-wave contribution persists across the transition, while the vortex contribution is what changes the infrared behavior of the full correlator.

\paragraph{Correlation length above $T_c$:}\label{sec:corr_length}
To quantify the onset of disorder we fit the full correlator above $b_c$ to the form
\be
G_2(r) = A\, r^{-b^2}\, e^{-r/\xi} ~,
\label{eq:G2_xi_fit}
\ee
which captures the power law envelope from the spin-wave sector and the exponential decay induced by vortex proliferation.
The exponent $b^2$ is fixed by the analytic result~\eref{eq:eta_b2}; $A$ and $\xi$ are free parameters determined by nonlinear least-squares in the window $r\in[2,30]$.

\begin{figure}[t]
\centering
\includegraphics[width=0.85\textwidth]{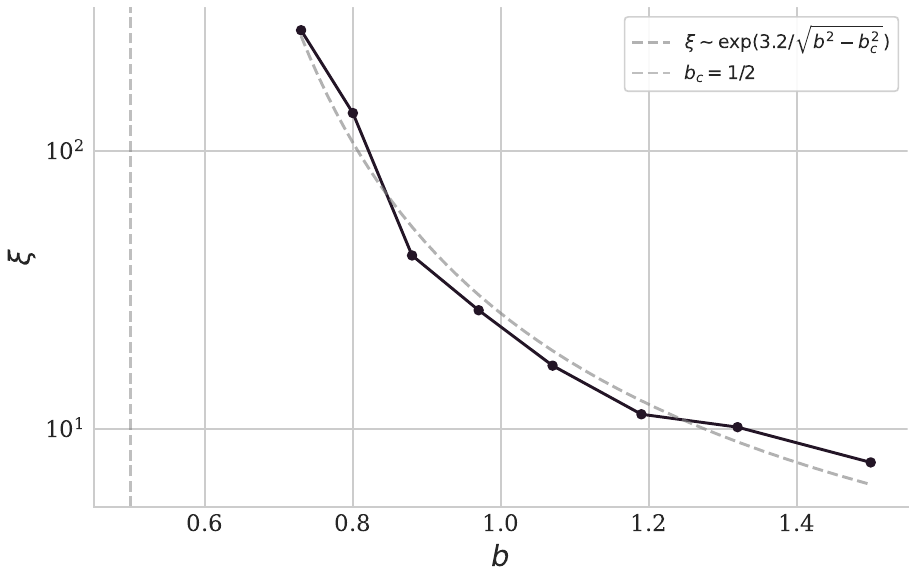}
\caption{Correlation length $\xi$ extracted from the full two-point
function via~\eref{eq:G2_xi_fit}.
$R^2$ values of the per-point fit~\eref{eq:G2_xi_fit} are annotated.
Dashed curve: BKT essential singularity fit~\eref{eq:xi_fit} ($R^2=0.98$).
\label{fig:corr_length}}
\end{figure}

Figure~\ref{fig:corr_length} shows the extracted correlation length $\xi(b)$ for $b>b_c$.
The per-point fits of~\eref{eq:G2_xi_fit} to $G_2(r)$ are excellent, with $R^2>0.97$ at every $b$.
The $b$-dependence of the extracted $\xi$ is well described by the BKT essential singularity
\be
\xi(b) \sim \exp\!\bigg(\frac{c}{\sqrt{b^2-b_c^2}}\bigg) ~,
\label{eq:xi_fit}
\ee
with $c\approx 3.17$ ($R^2=0.98$).
For a standard second-order transition one would instead expect a power law divergence $\xi\sim |T-T_c|^{-\nu}$; the best power law fit gives $\nu\approx 2.5$ with a worse $R^2=0.95$.
The essential singularity~\eref{eq:xi_fit} diverges faster than any power and is the hallmark of the BKT mechanism, consistent with the RG flow~\eref{eq:KT_RG}.
As $b\to b_c^+$, $\xi$ grows rapidly and eventually exceeds the system size, so the leftmost reliable points ($b\gtrsim 0.7$) have $\xi \gg L$.

\begin{figure}[t]
\centering
\includegraphics[width=0.7\textwidth]{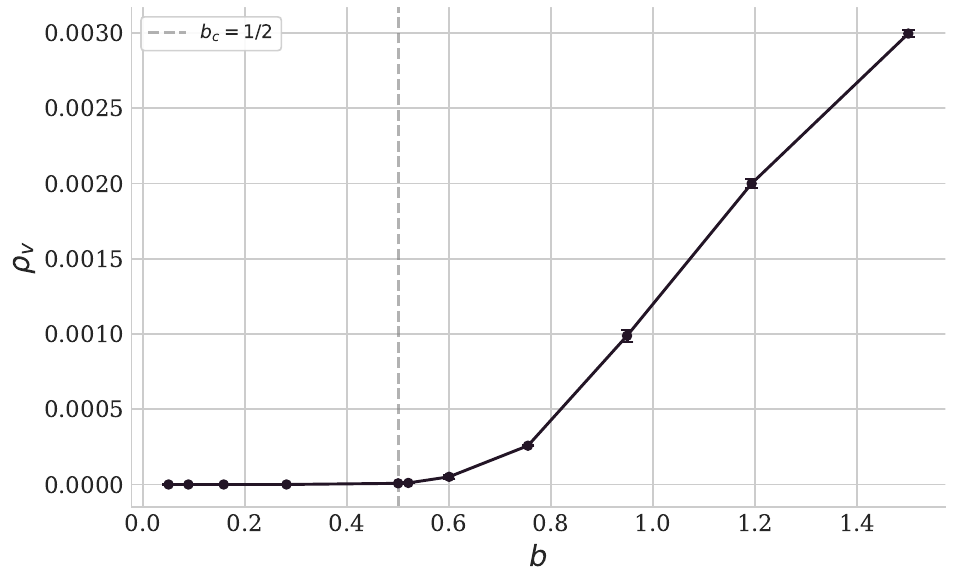}
\caption{Vortex density $\rho_v$ vs.\ $b$.
The vertical dashed line marks $b_c=1/2$.
Below $T_c$ the vortex density is consistent with zero; above $T_c$ it rises steeply as vortices unbind.
\label{fig:vortex_density}}
\end{figure}

\paragraph{Vortex density:}
A direct measure of vortex proliferation is the vortex density $\rho_v$, extracted from plaquette winding of the full field $\theta$.
Figure~\ref{fig:vortex_density} shows $\rho_v(b)$ across the transition.

Below $b_c$, vortices are bound into tightly correlated neutral pairs, and plaquette winding detects essentially none.
Above $b_c$, the density rises steeply as the energy--entropy balance tips in favor of free vortices.

\begin{figure}[t]
\centering
\includegraphics[width=0.92\textwidth]{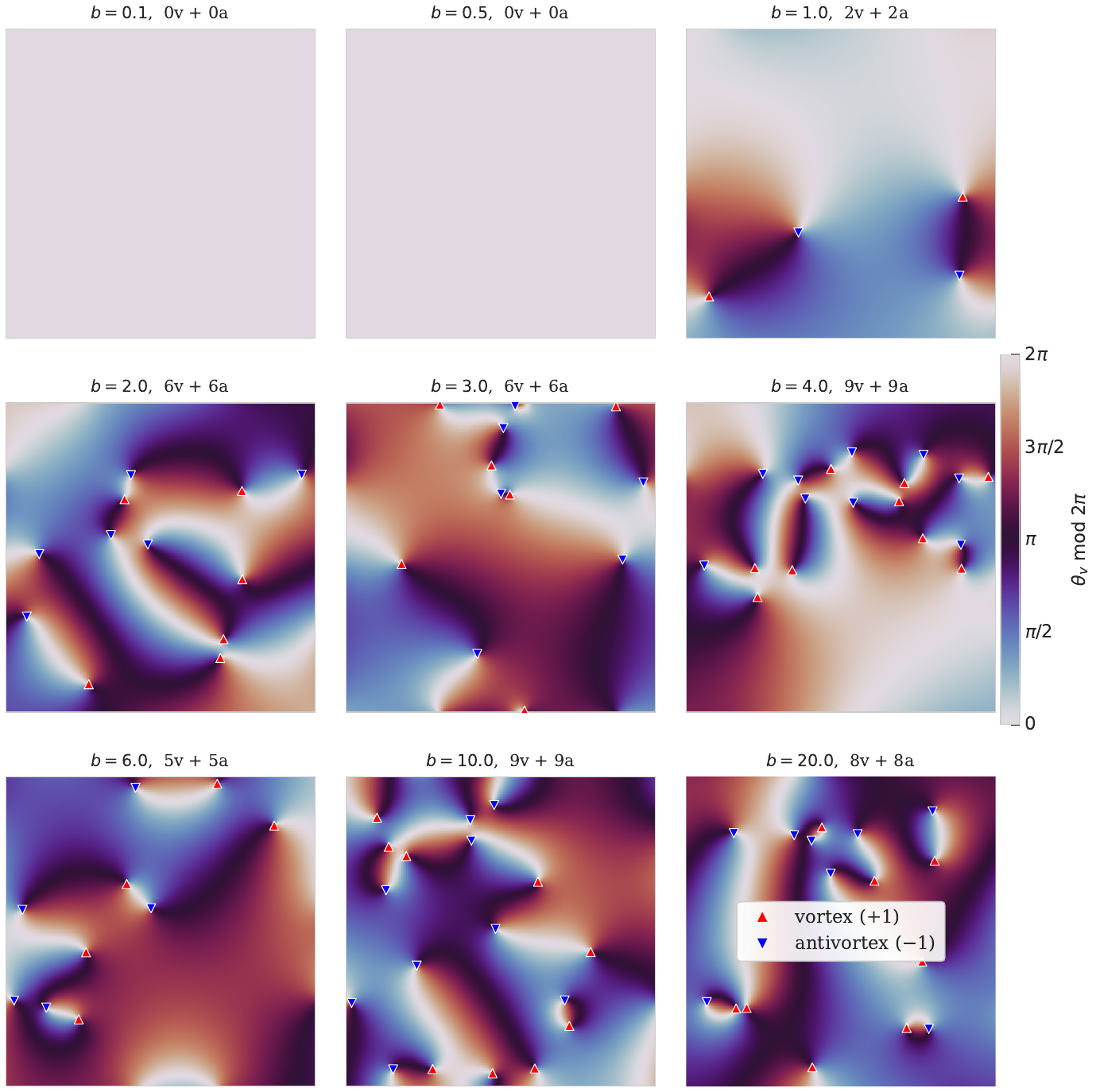}
\caption{Vortex angle field $\theta_v(x)\;\text{mod}\;2\pi$
for nine values of~$b$.
Red (blue) markers denote vortices (antivortices).
At small $b$ the field is featureless;
as $b$ increases past $b_c=1/2$, vortex--antivortex pairs appear and
eventually form a dense plasma.
\label{fig:vortex_map}}
\end{figure}

\paragraph{Vortex configuration maps:}
Figure~\ref{fig:vortex_map} shows the vortex angle field $\theta_v(x)\;\text{mod}\;2\pi$ on a high-resolution ($256\times 256$) grid for nine values of~$b$ spanning the transition.
Vortices (red triangles) and antivortices (blue triangles) are marked.

At small $b$ (large $K_0$), the Coulomb gas coupling is strong and the fugacity suppresses vortex creation: the field is smooth.
As $b$ increases past $b_c$, widely separated pairs appear.
At $b\gg b_c$ the system is a dense vortex plasma, consistent with the disordered phase.

\paragraph{Vortex--antivortex pair correlation:}
The pair correlation function $g_{+-}(r)$ measures the excess probability of finding an antivortex at distance~$r$ from a vortex, relative to an uncorrelated gas:
\be
g_{+-}(r) = \frac{h_{+-}(r)}{N_+\,\rho_-\,A_\text{shell}(r)} ~,
\label{eq:pair_corr_def}
\ee
where $h_{+-}(r)$ is the histogram of vortex--antivortex pair distances
in the radial bin at~$r$, $\rho_- = N_-/L^2$ is the mean antivortex density,
and $A_\text{shell}(r)=\pi(r_\text{out}^2 - r_\text{in}^2)$ is the annular shell area.
For a spatially uncorrelated gas, $g_{+-}(r)\to 1$.

\begin{figure}[t]
\centering
\includegraphics[width=0.75\textwidth]{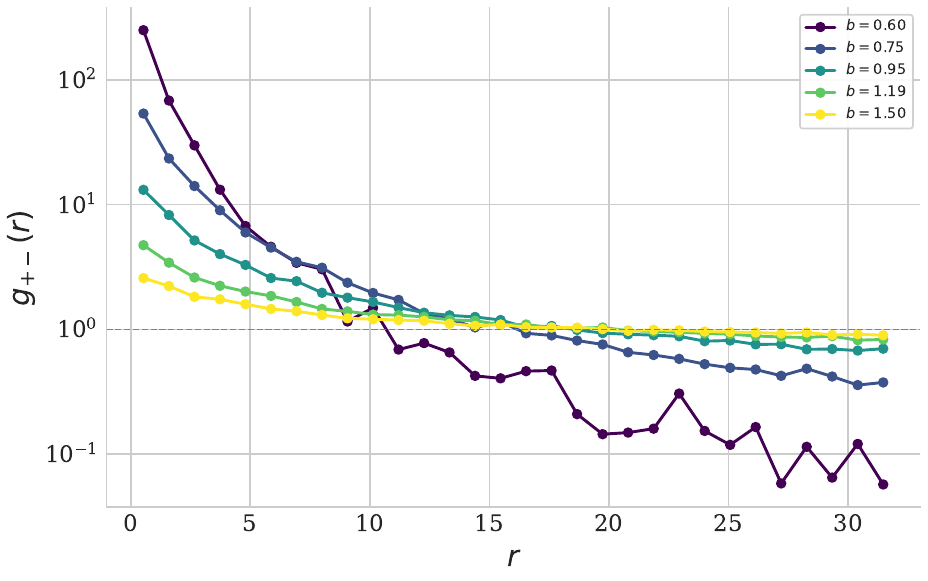}
\caption{Vortex--antivortex pair correlation $g_{+-}(r)$. At large $T$,  $b\to\infty$ and $g_{+ -}=1$ 
for all $r$, since the unbinding makes it an uncorrelated Coulomb gas (dashed line). At lower temperatures, we expect excess (deficit) probability at low (large) $r$, which is reflected in the indigo ($b=0.60$) and blue ($b=0.75$) curves, due to the logarithmic potential in $P_\text{vort}$.
The noisy statistics is due to a paucity of vortices.
\label{fig:pair_corr}}
\end{figure}

Figure~\ref{fig:pair_corr} shows $g_{+-}(r)$ for several values $b\ge 0.6$ (above $T_c$).
At all~$b$, $g_{+-}$ is strongly enhanced at small~$r$, reflecting the remnant Coulomb attraction between opposite charges.
Well above $T_c$ ($b \ge 0.95$), the curves settle to $g_{+-}\approx 1$ at large~$r$, confirming the approach to an uncorrelated plasma.
As $b$ increases, the binding becomes weaker and the enhancement at short distances diminishes, consistent with the transition from tightly bound pairs (near $T_c$) to a weakly interacting gas (high~$T$).

We should note that the two lowest curves ($b=0.60$ in indigo and $0.75$ in blue) in Figure~\ref{fig:pair_corr} are noisy and fall below unity at large~$r$: just above $b_c$ the vortex density is very low (\textit{cf.}~Figure~\ref{fig:vortex_density}), so few vortex--antivortex pairs contribute to the histogram and the statistical uncertainties are large.
Since the vortex gas is still extremely dilute just above $b_c$, $g_{+-}(r)$ is averaged only over configurations containing at least one vortex--antivortex pair, and in this regime the retained samples are typically dominated by a single bound pair.
The strong enhancement of short separations therefore comes at the expense of a depletion at larger separations, so $g_{+-}(r)$ can fall below the uncorrelated baseline $g_{+-}=1$ at large $r$.
The effect is further amplified by poor statistics in the outer radial bins, where long-distance pairs are rare.
Thus the large-$r$ undershoot is a dilute sample finite volume effect rather than evidence of a repulsive tail.

\paragraph{Helicity modulus from the power spectrum:}
We write the one-particle irreducible effective action in momentum space as
\be
S_\text{eff} = \frac12 \sum_k \Gamma^{(2)}(k) |\theta_k|^2 + \ldots ~.
\ee
Because a uniform shift of $\theta$ by a constant costs no energy, the zero-mode is free, and $\Gamma^{(2)}(0) = 0$.
If the long-distance theory is local and isotropic, then for small $k$, the one-particle irreducible two-point vertex function is
\be
\Gamma^{(2)}(k) = k^2 K_R + c_4 k^4 + \ldots ~.
\ee
The Green's function is the inverse of this quadratic kernel:
\be
G_\theta(k) = \langle |\theta_k|^2 \rangle \sim \frac{1}{\Gamma^{(2)}(k)} ~.
\ee
So
\be
\frac{1}{k^2 G_\theta(k)} = \frac{\Gamma^{(2)}(k)}{k^2} = K_R + c_4 k^2 + \ldots ~.
\ee
Taking the limit $k\to 0$, we strip off the higher order terms and obtain the renormalized stiffness from the inverse propagator:
\be
K_R = \lim_{k\to 0}\ \frac{1}{k^2 G_\theta(k)} ~.
\ee
Using~\eref{eq:ups}, we similarly define
\be
\Upsilon_R = T \lim_{k\to 0}\ \frac{1}{k^2 G_\theta(k)} ~.
\ee
This is our spectrum derived estimate of the helicity modulus.\footnote{
To be precise, in the numerical implementation, we extract the renormalized coupling $K_R$ from the momentum space correlator of the full field~$\theta$:
$$
K_R = \frac{L^4}{4\pi^2\,|n|^2\,\big\langle|\widetilde\theta(n)|^2\big\rangle} ~,
\label{eq:KR_def}
$$
where $\widetilde\theta(n)$ is the discrete Fourier transform of~$\theta$ at integer mode~$n$.
If the infrared theory is Gaussian with stiffness~$K_R$, then $\langle|\widetilde\theta(n)|^2\rangle = L^4/(4\pi^2 K_R |n|^2)$, so the previous expression inverts the propagator.
We average over the four lowest nonzero-modes ($|n|^2=1$) and multiply by $T$ to obtain the helicity.
In other words, the NN-FT implementation computes $K_R$ spectrally and infers $\Upsilon_R$ from this.}
Because it is extracted from the momentum-space correlator of the full field $\theta=b\,\theta_\text{sw}+\theta_v$, the same mixed ensemble that determines the full correlator $G_2(r)$ enters non-trivially in the determination of $\Upsilon_R$; the helicity modulus is therefore not inferred from the spin-wave sector alone, but is sensitive to vortex screening. This should coincide with the true helicity modulus in the infrared Gaussian regime, but it is not a direct twist response measurement as in the XY model.

\begin{figure}[t]
\centering
\includegraphics[width=0.85\textwidth]{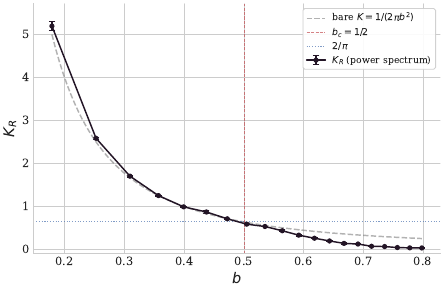}
\caption{Renormalized stiffness $K_R$ vs.\ $b$.
Gray dashed: bare coupling $K_0=1/(2\pi b^2)$.
Blue dotted: universal value $2/\pi$.
Red dashed: $b_c=1/2$.
Below $b_c$, $K_R\approx K_0$;
above $b_c$, vortex screening drives $K_R$ sharply downward.
\label{fig:stiffness}}
\end{figure}

Figure~\ref{fig:stiffness} shows $K_R$ as a function of~$b$, compared to the bare coupling $K_0=1/(2\pi b^2)$ and the universal value $K_R=2/\pi$.
Below $b_c$ the renormalized stiffness closely tracks the bare value: vortices are rare and their screening effect is small.
Above $b_c$, $K_R$ drops sharply as proliferating vortices screen the stiffness.

\begin{figure}[t]
\centering
\includegraphics[width=0.85\textwidth]{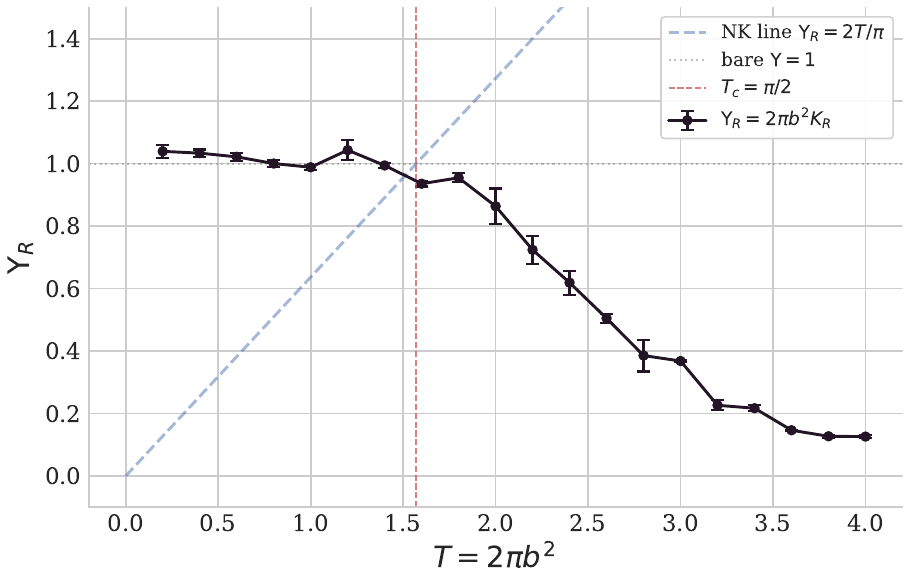}
\caption{Helicity modulus $\Upsilon_R = 2\pi b^2 K_R$ vs.\ temperature $T=2\pi b^2$.
Blue dashed: Nelson--Kosterlitz line $\Upsilon_R = 2T/\pi$.
Red dashed: $T_c = \pi/2$.
The data cross the NK line near $T_c$ as predicted,
and $\Upsilon_R$ collapses toward zero in the disordered phase.
\label{fig:helicity}}
\end{figure}

Figure~\ref{fig:helicity} translates these data into the helicity modulus $\Upsilon_R = T K_R$, which in the normalization $\rho_{s,0}=1$ becomes $\Upsilon_R = 2\pi b^2 K_R$, plotted against temperature $T=2\pi b^2$.
The Nelson--Kosterlitz prediction is that $\Upsilon_R$ meets the line $\Upsilon_R = 2T/\pi$ at $T_c$ with a universal jump~\eref{eq:universal_jump}.

The measured $\Upsilon_R(T)$ crosses the NK line near $T_c=\pi/2$, consistent with the universal jump.
Above $T_c$, the helicity modulus drops below the NK line and tends to zero, as expected for the vortex-plasma phase.
%

\subsection{Summary}\label{sec:bkt_summary}
Let us review the logic of our computation.
The radius of the circle $b$ fixes the bare input $K_0$.
The vortex gas and full field dynamics renormalize this to an infrared $K_R$.
The NN-FT implementation estimates $K_R$ from the spectrum of the full field.
The signal for different physics across the BKT transition is evident from the behavior of the two-point function in the different phases.
The renormalized helicity modulus, a dimensionful quantity, is defined as $\Upsilon_\text{R}=T K_R$.
A non-zero $\Upsilon_R$ means that the system can support persistent currents.

At low temperatures, vortices and anti-vortices are bound tightly in pairs.
A bound dipole pair has a localized phase field that does not disrupt the system's global ability to resist a phase twist.
Consequently, the macroscopic helicity modulus remains finite.
The system behaves as a superfluid with algebraically decaying correlation functions.

As temperature increases, the entropy gained by unbinding a vortex pair eventually overcomes the logarithmic energy cost of separating a single vortex pair.
The pairs unbind into a plasma of free vortices.
The full two-point function exhibits exponential decay with a finite correlation length.

Table~\ref{tab:summary} collects the observable signatures in the three
regimes of the NN-FT BKT construction.
\begin{table}[t]
\centering
\begin{tabular}{lccc}
\hline
Observable & $b < b_c$ (ordered) & $b=b_c$ (critical) & $b>b_c$ (disordered) \\
\hline
$\eta_\text{sw}$ & $b^2 < 1/4$ & $1/4$ & $b^2>1/4$ \\
$G_2(r)$ & $\sim r^{-b^2}$ & $\sim r^{-1/4}$ & $\sim r^{-b^2} e^{-r/\xi}$ \\
$\xi$ & $\infty$ & $\infty$ & finite, $\sim e^{c/\sqrt{b^2-b_c^2}}$ \\
$\rho_v$ & $\approx 0$ & onset & $>0$, increasing \\
$K_R$ & $\approx K_0 = 1/(2\pi b^2)$ & $2/\pi$ & $\ll 1/(2\pi b^2)$ \\
$\Upsilon_R$ & above NK line & on NK line & below NK line \\
\hline
\end{tabular}
\caption{Observable signatures across the BKT transition in NN-FT.
\label{tab:summary}}
\end{table}

The NN-FT mixed ensemble, comprised of Gaussian RFF spin-waves augmented by an explicit Coulomb gas vortex sector, reproduces the main qualitative signatures of BKT behavior in this numerical setup and yields several observables that are consistent with the expected transition near $b_c=1/2$ ($T_c=\pi/2$):
\begin{itemize}\itemsep0pt
\item The spin-wave correlator realizes the critical line $\eta=b^2$ with high precision.
\item The correlation length extracted from the full correlator diverges as $b\to b_c^+$ with the BKT essential singularity.
\item The vortex density vanishes below $T_c$ and rises steeply above.
\item The pair correlation $g_{+-}(r)$ shows the remnant Coulomb binding.
\item The helicity modulus crosses the universal Nelson--Kosterlitz line at the predicted critical temperature.
\end{itemize}
The vortex sector is essential: without it, the spin-wave-only theory remains on the Gaussian critical line at all~$b$, with no mechanism for disorder.
This analysis demonstrates that NN-FT can access topologically non-trivial physics when the continuous network architecture is augmented with discrete topological sector labels.

\section{T-duality}\label{sec:td}

We begin by establishing conventions for the bosonic string worldsheet sigma model in general background fields and by summarizing T-duality as an exact equivalence that relates backgrounds with a $U(1)$ isometry.
Let $(\Sigma,h_{ab})$ be the worldsheet with local coordinates $\sigma^a$ ($a=0,1$), metric $h_{ab}$, and scalar curvature $R^{(2)}[h]$.
We work in Euclidean signature on the worldsheet.
The embedding fields
\be
X^\mu(\sigma):\Sigma\to \mathcal{M} ~,
\ee
map the worldsheet into a $D$-dimensional target space with spacetime coordinates $X^\mu$ ($\mu=0,\dots,D-1$).
In the presence of a target space metric $G_{\mu\nu}(X)$, Kalb--Ramond $2$-form $B_{\mu\nu}(X)$, and dilaton $\Phi(X)$, the Euclidean sigma model action is
\be
\hspace{-5pt} S[X;h] = \frac{1}{4\pi\alpha'}\int_{\Sigma} d^2\sigma\,\sqrt{h}\,
\Big[
h^{ab}\,G_{\mu\nu}(X)\,\partial_a X^\mu \partial_b X^\nu
+ i\,\epsilon^{ab}\,B_{\mu\nu}(X)\,\partial_a X^\mu \partial_b X^\nu
+ \alpha'\,\Phi(X)\,R^{(2)}[h]
\Big].
\label{eq:ws_sigma_model}
\ee
Here, $\epsilon^{ab}$ is the antisymmetric tensor density with $\epsilon^{01}=+1/\sqrt{h}$, so that $\sqrt{h}\,\epsilon^{ab}$ is a tensor.
The $B$-field is topological on the worldsheet (it does not depend on $h_{ab}$), while the dilaton couples to the Euler characteristic via $\int\sqrt{h}\,R^{(2)}=4\pi\chi(\Sigma)$ when $\Phi$ is constant.

T-duality is most transparently stated when the target admits a $U(1)$ isometry. Choose adapted coordinates
\be
X^\mu=(y,x^i)
\ee
such that $y\sim y+2\pi$ parametrizes the isometry direction and all background fields are independent of $y$:
\be
\partial_y G_{\mu\nu}=\partial_y B_{\mu\nu}=\partial_y \Phi = 0 ~.
\ee
In this setting, the worldsheet theory~\eref{eq:ws_sigma_model} is equivalent to a dual sigma model in which $y$ is replaced by a dual coordinate $\tilde y$ and the background fields transform by the Buscher rules~\cite{Buscher:1987sk,Buscher:1987qj}.
Conceptually, one derives these rules by gauging the shift symmetry $y\to y+\text{constant}$ on the worldsheet, constraining the gauge field to be flat with a Lagrange multiplier $\tilde y$, and then integrating out either $\tilde y$ (recovering the original model) or the gauge field (producing the dual).
Equivalently, in the operator formalism for a free circle compactification, T-duality acts as a right mover parity on the compact boson,
\be
y_L(z)\ \mapsto\ y_L(z) ~,\qquad y_R(\bar z)\ \mapsto\ -y_R(\bar z) ~,
\ee
which exchanges Kaluza--Klein momentum and winding.

If we write the original background fields in components along the isometry direction $y$ and transverse directions $x^i$, the T-dual background $(\widetilde G,\widetilde B,\widetilde\Phi)$ is
\begin{align}
\widetilde G_{yy} &= \frac{1}{G_{yy}} ~, &
\widetilde G_{yi} &= \frac{B_{yi}}{G_{yy}} ~, &
\widetilde B_{yi} &= \frac{G_{yi}}{G_{yy}} ~, \label{eq:buscher1}\\[4pt]
\widetilde G_{ij} &= G_{ij}-\frac{G_{yi}G_{yj}-B_{yi}B_{yj}}{G_{yy}} ~, &
\widetilde B_{ij} &= B_{ij}-\frac{G_{yi}B_{yj}-B_{yi}G_{yj}}{G_{yy}} ~, \label{eq:buscher2}\\[4pt]
\widetilde \Phi &= \Phi - \frac{1}{2}\log G_{yy} ~.
\label{eq:buscher_dilaton}
\end{align}
These formulae should be interpreted as relations between sigma model couplings (string frame fields), and the dilaton shift~\eref{eq:buscher_dilaton} is required so that the functional measure (equivalently, the one-loop effective action) matches between the two descriptions.

For a constant background with
\be
ds^2 = R^2\,dy^2 + \delta_{ij}\,dx^i dx^j~, \qquad B=0 ~,\qquad \Phi=\Phi_0 ~,
\label{eq:simple_circle}
\ee
we have $G_{yy}=R^2$, so~\eref{eq:buscher1}--\eref{eq:buscher_dilaton} give
\be
\widetilde{ds}^2 = \frac{1}{R^2}\,d\tilde y^{\,2}+\delta_{ij}\,dx^i dx^j ~, \qquad \widetilde B=0 ~, \qquad \widetilde\Phi=\Phi_0-\log R ~.
\ee
Restoring the conventional radius $\widetilde R$ defined so that the dual circle has circumference $2\pi \widetilde R$ (\textit{i.e.}, rescaling $\tilde y$ to a $2\pi$-periodic coordinate), one obtains the familiar relation $\widetilde R = \alpha'/R$, together with the corresponding dilaton shift that keeps the effective string coupling invariant in the lower-dimensional theory.
In the free closed string spectrum, this is realized as the exchange of momentum and winding numbers, $n \leftrightarrow w$, under $R\leftrightarrow \alpha'/R$.

T-duality is an exact equivalence of worldsheet CFTs: when a background has a $U(1)$ isometry, one may trade the coordinate along the isometry for a dual coordinate, while $(G,B,\Phi)$ transform by~\eref{eq:buscher1}--\eref{eq:buscher_dilaton}.
The metric and $B$-field mix, and the dilaton shifts so that the quantum theory (including the path integral measure) is preserved.
This physics is well explained in~\cite{Polchinski:1998rq}.

\subsection{NN-FT realization of worldsheet sigma model}
To realize the bosonic string worldsheet embedding fields $X^\mu(z,\bar z)$ ($\mu=1,\dots,D$) as an NN-FT, one chooses the network input space to be the Euclidean worldsheet $\Sigma \simeq \mathbb{R}^2$ (or patches thereof), with complex coordinate $z=x^1+i x^2$, and the outputs to be the target space coordinates $X^\mu$.
A particularly transparent construction uses random Fourier feature (cosine) networks with a momentum space prior engineered so that the induced kernel is logarithmic.
Concretely, one takes an architecture of the schematic form 
\be
X^\mu(z,\bar z) \sim\frac{1}{\sqrt{N}}\sum_{i=1}^N \frac{a_i^\mu}{|W_i|} \cos\Big(\tfrac12(W_i z+\bar W_i \bar z)+c_i\Big) ~,
\label{eq:cos_architecture_schematic}
\ee
with i.i.d.\ parameters $a_i^\mu$ (Gaussian), phases $c_i$ (uniform), and complex ``frequencies'' $W_i$ drawn from an annulus in the complex plane with inner/outer radii $(\epsilon,\Lambda)$ that serve as IR/UV regulators.
In the $N\to\infty$ limit, one finds
\be
\langle X^\mu(z,\bar z)\,X^\nu(w,\bar w)\rangle = -\alpha' \delta^{\mu\nu}\log|z-w| + \text{(cutoff-dependent constant)} ~.
\label{eq:log_kernel}
\ee
Taking $1/\Lambda\ll |z-w|\ll 1/\epsilon$, we obtain the standard free boson worldsheet propagator; higher connected correlators vanish.
The physical parameter $\alpha'$ (string length squared) is controlled by the variance of the output weights (equivalently, by hyperparameters in $P(\theta)$); the string tension is then $T\sim 1/\alpha'$.

Two further standard ingredients also have natural NN-FT interpretations:
\begin{itemize}
\item \textbf{zero-modes:}
Adding a Gaussian ``bias'' (a constant random variable $X_0^\mu$) implements the $X^\mu$ zero-mode. In the flat prior limit for $X_0^\mu$, target space translation invariance implies momentum conservation in correlators of vertex operators~\cite{Frank:2026bui}.
\item \textbf{Target space symmetries:}
Worldsheet Euclidean invariance can be made manifest by rotationally invariant priors for the $W_i$, while $SO(D)$ invariance in target space is reflected by identical Gaussian priors for each $a_i^\mu$ component.
\end{itemize}

String theory demands more than global conformal invariance. 
We require that the gauge fixed worldsheet theory is a local two-dimensional CFT with a holomorphic stress tensor $T(z)$ generating the Virasoro algebra.
Generic generalized free fields lack a local conserved stress tensor obeying conformal Ward identities, which obstructs a direct NN-FT realization of Virasoro symmetry.
A key development is the construction of an NN-FT whose infinite width limit admits a local stress tensor.
$T(z)$ arises in~\cite{Robinson:2025ybg} from a log-kernel network $\varphi$
with a scale invariant spectral density $p(k)\propto |k|^{-2}$ chosen precisely so that $\langle \varphi(z)\varphi(0)\rangle\sim -\log|z|^2$. At infinite width the associated Gaussian process of that network is equivalent to that of~\eref{eq:cos_architecture_schematic}.\footnote{The two constructions use complementary conventions for distributing the spectral weight. In~\eref{eq:cos_architecture_schematic}, following~\cite{Frank:2026bui}, the $1/|W_i|$ factor is built into the architecture while the frequencies $W_i$ are drawn from a uniform (flat) prior on an annulus; the kernel becomes logarithmic because the architectural weighting supplies the needed $|k|^{-2}$ suppression mode by mode. In~\cite{Robinson:2025ybg}, the architecture is unweighted but the frequencies are instead drawn from a scale invariant spectral density $p(k)\propto |k|^{-2}$, so that the same $|k|^{-2}$ factor enters through the prior rather than the architecture. Both prescriptions yield the same infinite width Gaussian process and hence the same log kernel~\eref{eq:log_kernel}; they differ only in whether the spectral weight is carried by the network map or by the parameter density as noted in~\cite{Frank:2026bui}.}
 With this kernel, the usual holomorphic composite
\be
T(z) \propto\ :\big(\partial\varphi(z)\big)^2:
\label{eq:stress_tensor}
\ee
is well defined and its correlators reproduce the Virasoro algebra.
In particular, the construction yields $c=1$ and reproduces the expected scaling dimensions of vertex operators, and it admits extensions to free fermions and super-Virasoro symmetry~\cite{Robinson:2025ybg}.
These ingredients supply precisely the local conformal structure needed for worldsheet string theory.

Gauge fixing the Polyakov path integral introduces the reparametrization ghosts $b,c$, whose action is that of a free first order system and whose central charge is $c_{bc}=-26$.
The $bc$ system is obtained using Grassmann valued neural networks~\cite{Frank:2026bui}.
One chooses architectures for $b(z,\bar z)$ and $c(z,\bar z)$ built from complex Fourier modes with Grassmann coefficients, together with a Berezin integrable Gaussian prior on the Grassmann parameters.
The resulting two-point functions reproduce the standard $bc$ propagators (up to regulator effects), and higher connected correlators again vanish in the infinite width limit.

Putting the pieces together, the NN-FT ensemble reproduces the gauge fixed bosonic worldsheet theory
\be
S_\text{ws} = \frac{1}{2\pi\alpha'}\int d^2 z\, \partial X^\mu \bar\partial X_\mu
+\frac{1}{2\pi}\int d^2 z\, \big(b\,\bar\partial c+\bar b\,\partial \bar c\big) ~,
\label{eq:gauge_fixed_polyakov}
\ee
now with the functional integral realized as an average over network parameter space.
In this form, the usual critical dimension logic is expected to carry over: each free boson contributes $c=1$, so conformal anomaly cancellation suggests $D=26$ once the NN-FT realization of the ghost central charge is tracked carefully.

With $X^\mu$ realized as a Gaussian process with covariance~\eref{eq:log_kernel}, correlators of normal ordered vertex operators
\be
V_k(z,\bar z)=\ :e^{i k\cdot X(z,\bar z)}:
\label{eq:vertex_ops}
\ee
reduce to the standard Koba--Nielsen form by Wick's theorem.
In the NN-FT picture, the same object is computed by first evaluating $V_k$ on the network output and then averaging over $\theta$ with the prescribed prior.
The resulting parameter space correlators, after integrating over insertion points and dividing by the residual conformal Killing volume, reproduce the classic tree level bosonic string amplitudes (Veneziano and Virasoro-Shapiro)~\cite{Frank:2026bui}.
The conceptual novelty is not the amplitude evaluation, but the re-expression of the worldsheet path integral as an integral over neural network parameter space.

\subsection{Evidence for T-duality in NN-FT}\label{sec:tdual_evidence}

The compact boson provides the cleanest benchmark of T-duality in the present framework.
The oscillator sector is supplied by the Gaussian NN-FT sampler~\eref{eq:cos_architecture_schematic}, whereas the compact zero-modes are represented explicitly as discrete momentum/winding latents. 
Accordingly, the compact boson ensemble is of the mixed form introduced in~\eref{eq:param_expectation_mixed},
\be
\langle \mathcal O\rangle_R
=
\sum_{n,w\in\mathbb Z}\int d\Theta_\text{osc}\;
P_\text{osc}(\Theta_\text{osc})\,P_R(n,w)\,
\mathcal O[X_{\Theta_\text{osc};n,w}] ~,
\label{eq:tdual_mixed_expectation}
\ee
with $P_R(n,w)$ the standard compact lattice weight.
No claim is made here that winding sectors emerge dynamically from a generic neural prior.
The aim is narrower: given a finite width oscillator sampler and an explicit compact lattice sector, do the combined ensembles satisfy the standard T-duality identities for states, correlators, and constant toroidal backgrounds?
The answer is affirmative.

In the exact circle CFT, T-duality acts by
\be
R\ \longrightarrow\ \widetilde R=\frac{\alpha'}{R} ~,\qquad n\longleftrightarrow w ~,
\ee
or, equivalently, by leaving $p_L$ invariant and sending $p_R\to -p_R$.
This makes the compact boson especially well suited to a numerical NN-FT test: one can compare the same random-feature realization of the oscillator field against two different compact zero-mode assignments, so any discrepancy with duality can be traced directly to finite width effects.

Our strategy is therefore to compare Monte Carlo ensembles at radius $R$ and at the dual radius $\widetilde R$, with the NN-FT extended so that the compact sector carries both momentum and winding.
The evidence is organized in three increasingly stringent layers.
First, we test the exchange of momentum and winding in the sampled topological sectors themselves, including the induced transformation of $p_L$, $p_R$, the corresponding conformal weights, and the monodromy of the field around the spatial circle.
Second, we consider a two-torus in target space and ask whether the transformation of the compact background itself, viewed through the worldsheet couplings encoded by the sampler, matches the Buscher rules for the background.
Third, using a chiral finite mode version of the same construction, we compare the operator data on the two sides of the duality by checking vertex operator selection rules, two-point functions, and extracted scaling dimensions.
In this way the numerics test whether the NN-FT realizes T-duality as an equivalence of sampled ensembles and observables, rather than merely as a formal statement about the continuum circle CFT.

\paragraph{Exchange of momentum and winding:}
At the implementation level, this first test isolates the minimal way T-duality can appear in NN-FT. The oscillator sampler remains periodic and unchanged, while the explicit discrete zero-mode latents carry the momentum--winding data. Aside from the bookkeeping step of re-expressing the zero-mode position $x_0$ on the dual circle, the dual map acts non-trivially only on these compact charges. The lesson is that, once the compact lattice is supplied explicitly, a finite width NN-FT can realize the topological content of circle duality sample by sample: local fluctuations and global monodromy stay cleanly separated, and duality is implemented through the compact sector rather than having to emerge indirectly from the oscillator prior.

For this first one circle test it is convenient to work with the physical compact coordinate $X\sim X+2\pi R$. We place the worldsheet on a cylinder with coordinates $(\tau,\sigma)$, $\sigma\sim \sigma+2\pi$, and modify the cosine feature sampler~\eref{eq:cos_architecture_schematic} by leaving its oscillator sector unchanged and compactifying only the zero-mode. A single field draw is taken to be
\begin{align}
X(\tau,\sigma) &= X_\text{osc}(\tau,\sigma)+x_0+\frac{\alpha'}{2}p_L(\tau+\sigma)+\frac{\alpha'}{2}p_R(\tau-\sigma) \nonumber \\
&= X_\text{osc}(\tau,\sigma)+x_0+\alpha'\frac{n}{R}\,\tau + wR\,\sigma ~,
\label{eq:NN-FT_compact_boson_sample}
\end{align}
with $x_0$ uniform on the circle and
\be
p_L=\frac{n}{R}+\frac{wR}{\alpha'} ~, \qquad p_R=\frac{n}{R}-\frac{wR}{\alpha'} ~.
\label{eq:NN-FT_compact_boson_momenta}
\ee
For the Monte Carlo we work at vanishing real modular\footnote{Here $\tau_1,\tau_2$ denote the torus modular parameters, whereas $\tau$ continues to denote the cylinder time coordinate.} parameter, $\tau_1=0$, so the topological sector has a positive weight
\be
P_R(n,w)\propto \exp\!\left[-\pi\tau_2\left(\alpha'\frac{n^2}{R^2}+\frac{R^2}{\alpha'}w^2\right)\right] ~,
\label{eq:NN-FT_compact_boson_weight}
\ee
which we sample directly after truncating the lattice to $|n|,|w|\le 12$. Although the field draws below are written on a cylinder, this is precisely the standard $\tau_1=0$ torus weight for the compact lattice, so the sampled discrete sectors are the same ones that enter the continuum partition function.

Under $R\to \widetilde R=\alpha'/R$ and $n\leftrightarrow w$, one has $p_L\to p_L$ and $p_R\to -p_R$, so the zero-mode action is unchanged and the conformal weights $h=\tfrac{\alpha'}{4}p_L^2$ and $\bar h=\tfrac{\alpha'}{4}p_R^2$ are preserved as well.
The point of the numerical check is therefore not to rederive the duality from scratch, but to verify that the finite width sampler reproduces the standard compact boson map faithfully on actual field draws once the momentum--winding sector is supplied explicitly.

Numerically we set $\alpha'=1$, $\tau_2=0.9$, and, for the sweep and quoted Monte Carlo statistics, used $96$ random cosine features.
For each radius $R\in\{0.65,0.80,1.00,1.25,1.70,2.20\}$, we generated $10^6$ direct samples of the topological sector and $10^4$ paired field configurations in which the oscillator draw was held fixed while $(n,w)$ was exchanged across the dual radii.
The exact truncated lattice identity $P_R(n,w)=P_{\widetilde R}(w,n)$ is satisfied to machine precision throughout the sweep, with $\max|\Delta P|\le 2.2\times 10^{-16}$.

The sampled momentum and winding histograms agree at the corresponding Monte Carlo level. Concretely, from the $10^6$ sampled topological sectors at radius $R$ and at the dual radius $\widetilde R=\alpha'/R$, we constructed normalized empirical distributions $\hat P_R(n,w)$ and $\hat P_{\widetilde R}(n,w)$ and compared $\hat P_R(n,w)$ to the dual distribution after exchanging momentum and winding, $\hat P_{\widetilde R}(w,n)$. The discrepancy was quantified by the total variation distance,
\be
d_{\rm TV}
:=
\frac12\sum_{n,w}
\left|
\hat P_R(n,w)-\hat P_{\widetilde R}(w,n)
\right| ~,
\ee
which is the standard half-$\ell^1$ distance between two discrete probability distributions. Across the full scan we find
\be
d_{\rm TV}\in [2.9\times10^{-4},\,9.3\times10^{-4}] ~,
\ee
showing that, once the expected T-duality exchange $n\leftrightarrow w$ is made, the sampled topological-sector distributions at $R$ and $\widetilde R$ agree to within Monte Carlo error.

At the field level, the monodromy of the unwrapped field
\be
\Delta_\sigma X := X(\tau,2\pi)-X(\tau,0)=2\pi wR
\label{eq:NN-FT_compact_boson_monodromy}
\ee
is reproduced with errors of order $10^{-15}$, and the same holds after dualization with $w$ replaced by $n$ and $R$ by $\widetilde R$.
Here the check is performed before reducing back to the wrapped compact representative, so the winding formula holds literally on each draw. Because the oscillator contribution is periodic in $\sigma$, this monodromy isolates the topological sector directly.
We also checked the exchanged zero-mode slopes, $\partial_\tau X_\text{zm}=\alpha' n/R$ and $\partial_\tau \widetilde X_\text{zm}=\alpha' w/\widetilde R$, again finding agreement at the $10^{-15}$ level.
A representative run at $R=1.7$ ($\widetilde R\simeq 0.588$) gives exact $\max|\Delta P|=2.2\times 10^{-16}$, empirical total variation distance $6.7\times 10^{-4}$, maximal winding error $7.1\times 10^{-15}$, and maximal exchanged-slope error $1.1\times 10^{-15}$.
This is precisely the part of T-duality visible in the compact boson itself: the NN-FT sampler exchanges momentum and winding, preserves $p_L$, flips the sign of $p_R$, and leaves the conformal weights invariant.

\paragraph{Buscher rules:}
The next question is what part of the target space background the NN-FT actually encodes. In this implementation the covariance of the vector-valued oscillator features captures the local metric dependence of the propagating modes, while the constant metric and $B$-field also enter the compact sector through the lattice momenta, conformal weights, action, and sampling weights. The dilaton is not part of the sampler itself; instead, after dualization we verify separately that the lower-dimensional invariant $e^{-2\Phi}\sqrt{\det G}$ is preserved under $\widetilde\Phi=\Phi-\frac12\log G_{yy}$. The point of the test is therefore to ask whether a finite width sampler realizes T-duality as a transformation of sigma model couplings, not just as a reshuffling of discrete charges; if it does, then the architecture is correctly separating and recombining the local propagator data and the global topological data.

To test the background dependent part of the duality, namely the transformation of the sigma model couplings in~\eref{eq:buscher1}--\eref{eq:buscher_dilaton}, we pass to the minimal setting in which metric and Kalb--Ramond data can mix non-trivially: a two-torus with dimensionless coordinates $X^I=(y,x)$, $I=1,2$, each identified as $X^I\sim X^I+2\pi$, and with duality performed along $y$. In these conventions the radii and shape are carried entirely by the constant target space metric $G_{IJ}$, while the constant antisymmetric $B$-field enters only through the compact lattice. The cosine-feature architecture is then modified in two ways. First, the oscillator part becomes vector valued, with the same random frequencies and phases as before but with feature amplitudes promoted to vectors $a_r^I$. Writing $z=e^{\tau+i\sigma}$, we take
\be
X_\text{osc}^I(z,\bar z)
=
\frac{C}{\sqrt{N}}\sum_{r=1}^N \frac{a_r^I}{|W_r|}
\cos\!\big(\Re(W_r z)+c_r\big) ~,\qquad
\langle a_r^I a_s^J\rangle = \delta_{rs}\,\sigma_a^2\,G^{IJ} ~,
\ee
so that the induced Gaussian kernel is the matrix-valued free boson propagator appropriate to the constant-metric sigma model. Second, the compact zero-mode is promoted from a single pair $(n,w)$ to momentum and winding vectors $n_I,w^I\in\mathbb{Z}$,
\be
X^I(\tau,\sigma)=X_\text{osc}^I(\tau,\sigma)+x_0^I+\alpha' p^I \tau + w^I \sigma ~,\qquad
p^I=(G^{-1})^{IJ}\big(n_J+B_{JK}w^K\big) ~,
\ee
with
\be
p_L^I=p^I+\frac{w^I}{\alpha'} ~,\qquad
p_R^I=p^I-\frac{w^I}{\alpha'} ~,\qquad
h=\frac{\alpha'}{4}p_L^T G p_L ~,\qquad
\bar h=\frac{\alpha'}{4}p_R^T G p_R ~.
\ee
Because the background is constant, the $B$-field does not alter the local oscillator kernel; it enters entirely through the shifted compact lattice. At $\tau_1=0$ the topological sector is therefore sampled with positive weight
\be
P_{G,B}(n,w)\propto \exp\!\big[-2\pi\tau_2(h+\bar h)\big] ~,
\ee
while Buscher duality acts by~\eref{eq:buscher1}--\eref{eq:buscher_dilaton} on $(G,B,\Phi)$ and by
\be\label{eqn:buscher_charge_map}
(n_y,n_x;w^y,w^x)\ \longmapsto\ (w^y,n_x;n_y,w^x)
\ee
on the compact charges.

Numerically we set $\alpha'=1$, $\tau_2=0.5$, $\epsilon=0.5$, $\Lambda=6.0$, and used $128$ random cosine features. For a representative background
\be\label{eq:buscher_G_and_B}
G=\begin{pmatrix}2.89&0.45\\[2pt]0.45&1.69\end{pmatrix} ~,\qquad
B=\begin{pmatrix}0&0.38\\[2pt]-0.38&0\end{pmatrix} ~,
\ee
the Buscher transform along $y$ gives
\be
\widetilde G=\begin{pmatrix}
0.3460&0.1315\\[2pt]
0.1315&1.6699
\end{pmatrix} ~,\qquad
\widetilde B=\begin{pmatrix}
0&0.1557\\[2pt]
-0.1557&0
\end{pmatrix} ~.
\ee
We truncated the charge lattice to $|n_I|,|w^I|\le 5$ and first compared the exact discrete distributions on the two sides. The Buscher-mapped identity
\be
P_{G,B}(n_y,n_x;w^y,w^x)
=
P_{\widetilde G,\widetilde B}(w^y,n_x;n_y,w^x)
\ee
is satisfied to machine precision on the truncated lattice, with $\max|\Delta P|=4.2\times 10^{-17}$, $\max|\Delta S|=1.7\times 10^{-13}$, $\max|\Delta h|=3.6\times 10^{-14}$, and $\max|\Delta\bar h|=2.8\times 10^{-14}$. We then drew $6\times 10^4$ sectors from each ensemble and $2\times 10^3$ paired field configurations in which the oscillator latents were held fixed while the compact charges and background fields were dualized. The $(n_y,w^y)$ marginal histograms agree with the exact Buscher-transformed marginal at the expected Monte Carlo level, with total variation distances $2.6\times 10^{-3}$ for the original ensemble and $3.5\times 10^{-3}$ for the mapped dual ensemble.

At the field level, the topological monodromy again isolates the compact sector directly. Since $X^I\sim X^I+2\pi$, one has
\be
\Delta_\sigma X^I := X^I(\tau,2\pi)-X^I(\tau,0)=2\pi w^I ~,
\ee
and the maximal error in this relation is $3.6\times 10^{-15}$ in the original background and $8.9\times 10^{-15}$ after dualization. The direct Buscher exchange $n_y\to \widetilde w^y$ is satisfied with the same $8.9\times 10^{-15}$ accuracy. To check that the oscillator sector itself carries the transformed metric, we also measured the empirical covariance matrix from $5\times 10^3$ oscillator draws at a fixed worldsheet point; after normalizing by the trace, the resulting covariance agrees with the shapes of $G^{-1}$ and $\widetilde G^{-1}$ with errors $8.0\times 10^{-3}$ and $5.7\times 10^{-3}$, respectively. Finally, although the dilaton is checked separately from the sampler, including the constant dilaton shift $\widetilde\Phi=\Phi-\frac12\log G_{yy}$, the lower-dimensional combination $e^{-2\Phi}\sqrt{\det G}$ is preserved to floating point precision. Thus the finite width NN-FT sampler reproduces the Buscher transformation of the constant background data relevant to the sampled fields: the metric and $B$-field mix exactly as in the worldsheet sigma model, the compact charge lattice is mapped correctly, and the expected dilaton shift preserves the lower-dimensional invariant.

\paragraph{Vertex operators:}
State-level agreement does not yet guarantee the relevant operator data. We will now test the compact boson selection rules, two-point functions, fitted scaling dimensions, and paired-draw duality map more directly. The finite cutoff sampler is still organized in the same way as the compact boson CFT: chiral covariance controls the oscillator part, compact charges control the neutrality conditions, and the duality acts directly on operator labels together with $X_R\to -X_R$.

For operator-level checks it is convenient to replace the earlier nonchiral random-cosine sampler (\ref{eq:NN-FT_compact_boson_sample}) by an equivalent chiral finite-mode truncation on the cylinder, in which $X_L$, $X_R$, normal ordering, and zero-mode neutrality are explicit. This is a change of regulator/conventions, not of the underlying free compact boson theory being tested. We therefore specialize to the Euclidean cylinder with $\sigma\sim \sigma+2\pi$ and chiral coordinates
\(
u=\tau+\sigma
\),
\(
v=\tau-\sigma
\).
For the vacuum sector relevant to local correlators, let
\be
X(\tau,\sigma)=X_L(u)+X_R(v) ~,\quad
X_L(u)=\frac12(x_0+\tilde x_0)+X_L^{\rm osc}(u) ~,\quad
X_R(v)=\frac12(x_0-\tilde x_0)+X_R^{\rm osc}(v) ~,
\label{eq:compact_chiral_field}
\ee
where \(x_0\in[0,2\pi R)\) and \(\tilde x_0\in[0,2\pi\alpha'/R)\) are uniform zero-modes and the oscillator pieces are sampled with a finite chiral cosine-feature truncation
\be
X_{L/R}^{\rm osc}(\xi)=\sqrt{\frac{\alpha'}{2}}\sum_{m=1}^{M}
\frac{a_m^{L/R}\cos(m\xi)+b_m^{L/R}\sin(m\xi)}{\sqrt m} ~,\qquad M=64 ~,
\label{eq:compact_chiral_features}
\ee
with independent Gaussian parameters \(a_m^{L/R},b_m^{L/R}\).
The vertex operators are then
\be
V_{n,w}(u,v)=:\exp\!\big(i p_L X_L(u)+i p_R X_R(v)\big): ~,\qquad
p_L=\frac{n}{R}+\frac{wR}{\alpha'} ~,\qquad
p_R=\frac{n}{R}-\frac{wR}{\alpha'} ~,
\label{eq:compact_vertex_ops}
\ee
with conformal weights \(h=\alpha' p_L^2/4\) and \(\bar h=\alpha' p_R^2/4\). At finite mode cutoff the chiral covariance is
\be
C_M(\Delta)=\frac{\alpha'}{2}\sum_{m=1}^{M}\frac{\cos(m\Delta)}{m} ~,
\label{eq:compact_chiral_covariance}
\ee
so normal ordering is implemented by subtracting the self-contraction $C_M(0)$, \textit{i.e.}\ by multiplying each insertion by $\exp[\frac12(p_L^2+p_R^2)C_M(0)]$.
Averaging over the compact zero-modes enforces the usual neutrality conditions $\sum_i n_i=0$ and $\sum_i w_i=0$, and for normal-ordered correlators one obtains
\begin{align}
\Big\langle \prod_{i=1}^k V_{n_i,w_i}(u_i,v_i)\Big\rangle_M
&=
\delta_{\sum_i n_i,0}\,\delta_{\sum_i w_i,0} \nonumber \\
&\qquad \cdot \exp\!\Bigg[
-\sum_{i<j} p_{L,i}p_{L,j}\,C_M(u_i-u_j)
-\sum_{i<j} p_{R,i}p_{R,j}\,C_M(v_i-v_j)
\Bigg] ~.
\label{eq:compact_vertex_correlator}
\end{align}
In particular, for an equal-time pair one finds
\be
\Big\langle V_{n,w}(0,0)V_{-n,-w}(\Delta,-\Delta)\Big\rangle_M
=
\exp\!\Big[(p_L^2+p_R^2)\,C_M(\Delta)\Big]
\xrightarrow[M\to\infty]{}
\big(2\sin\tfrac{|\Delta|}{2}\big)^{-2(h+\bar h)} ~.
\label{eq:compact_vertex_two_point}
\ee

We evaluated these observables by direct Monte Carlo averaging over independent NN-FT draws. As a first check, the compact zero-modes enforce the expected selection rules numerically: the one-point functions of non-trivial operators are consistent with zero,
$
|\langle V_{1,0}\rangle|=2.8\times10^{-3}
$,
$
|\langle V_{0,1}\rangle|=3.3\times10^{-3}
$,
and
$
|\langle V_{1,1}\rangle|=1.4\times10^{-3}
$,
while a raw two-point function with insertions at $(u,v)=(0,0)$ and $(0.9,-0.9)$ gives
$
|\langle V_{1,0}V_{-1,0}\rangle|=0.1656
$
versus the finite $M$ prediction $0.1667$, whereas the non-neutral combinations
$
|\langle V_{1,0}V_{1,0}\rangle|=4.1\times10^{-4}
$
and
$
|\langle V_{1,0}V_{0,1}\rangle|=2.5\times10^{-3}
$
are strongly suppressed.
The absolute normalization of the un-normal-ordered correlator is cutoff dependent, but the near-vanishing of the non-neutral channels is the expected compact boson selection rule.

T-duality acts on the operator data by
\be
R\ \longrightarrow\ \widetilde R=\frac{\alpha'}{R} ~,\qquad
(n,w)\ \longrightarrow\ (w,n) ~,\qquad
X_R\ \longrightarrow\ -X_R ~,
\label{eq:compact_vertex_tduality}
\ee
so that $p_L$ is invariant and $p_R$ changes sign.
For $R=1.6$ and $\alpha'=1$, hence $\widetilde R=0.625$, the momentum operator $V_{1,0}$ at radius $R$ and the winding operator $V_{0,1}$ at radius $\widetilde R$ therefore have the same scaling dimension $h+\bar h = 25/128 \approx 0.1953$.
This was verified in a normal-ordered equal-time scan of~\eref{eq:compact_vertex_two_point} over $\Delta\in[0.4,2.4]$: the Monte Carlo curves agree with the corresponding finite $M$ predictions with maximal relative errors $2.4\%$ and $3.0\%$, respectively, and the fitted dimensions are
\be
\Delta_\text{fit}^{(1,0)}(R)=0.1996 ~,\qquad
\Delta_\text{fit}^{(0,1)}(\widetilde R)=0.1975 ~,
\label{eq:compact_vertex_fit}
\ee
both close to the exact value $25/128$.
Moreover, when original and dual draws are paired by exchanging the compact and dual zero-modes and sending $X_R\to -X_R$, the sample-by-sample identity
$
V_{n,w}^{(R)}(u,v)=V_{w,n}^{(\widetilde R)}(u,v)
$
holds to machine precision for the tested charges $(1,0)$, $(1,1)$, and $(2,1)$.

For larger charges and multipoint functions, direct sampling of the exponentials develops larger variance, but the oscillator sector at fixed finite mode cutoff is still Gaussian, so one can build a lower-variance estimator that targets the exact finite $M$ Gaussian prediction from Monte Carlo estimates of the chiral covariance together with the zero-mode selection rules.
Using this lower-variance estimator we found
\be
\langle V_{1,1}(0,0)V_{-1,-1}(0.4,-0.4)\rangle = 16.38\pm0.45
\ee
versus the finite $M$ prediction $16.30$,
\be
\langle V_{2,1}(0,0)V_{-2,-1}(0.4,-0.4)\rangle = 50.52\pm2.15
\ee
versus $49.37$,
and the mixed four-point function
\be
\langle V_{1,0}(0,0)V_{0,1}(0.7,-0.3)V_{-1,0}(1.4,-0.9)V_{0,-1}(2.1,-1.5)\rangle
=0.855\pm0.027
\ee
versus $0.836$; in each case the independently sampled dual correlator agreed within errors with both the original correlator and the shared T-dual prediction.
Taken together, these tests provide quantitative evidence that the compact boson NN-FT reproduces the standard T-duality map in the regime where the worldsheet theory is the free circle CFT: the ensemble enforces momentum/winding neutrality, reproduces the expected compact boson conformal weights, and yields matching momentum and winding correlators under $R\leftrightarrow \alpha'/R$.

\paragraph{Self-dual radius and enhanced current algebra:}
At the self-dual radius the test becomes substantially sharper, because the NN-FT must reproduce symmetry enhancement rather than merely invariance of the compact spectrum. We will now show that the implementation places the appropriate momentum-winding operators at dimension $1$ and organizes the lightest primaries into the expected $SU(2)_L\times SU(2)_R$ multiplets~\cite{Gross:1985fr,Narain:1985jj}. The representation theoretic part is checked algebraically from the self-dual charges, while the sampled ensemble is used for current two-point data and one representative current-primary channel. We note that the compact boson CFT at the self-dual radius is equivalent to the $SU(2)$ WZW model at level $k=1$~\cite{Witten:1983ar}, so the present construction can also be interpreted as an NN-FT realization of that theory and a numerical study of its observables.

More concretely, at the fixed point
\be
R=\widetilde R=\sqrt{\alpha'} ~,
\label{eq:selfdual_radius}
\ee
the compact boson develops the familiar enhanced
$SU(2)_L\times SU(2)_R$ current algebra.
In the conventions of~\eref{eq:compact_vertex_ops}, the charged currents are realized by
\be
J_L^\pm(u)=V_{\pm1,\pm1}(u,v) ~,\qquad
J_R^\pm(v)=V_{\pm1,\mp1}(u,v) ~,
\label{eq:selfdual_charged_currents}
\ee
while the Cartan currents are
\be
J_L^3(u)=\frac{i}{\sqrt{\alpha'}}\,\partial X_L(u) ~,\qquad
J_R^3(v)=\frac{i}{\sqrt{\alpha'}}\,\bar\partial X_R(v) ~.
\label{eq:selfdual_cartan_currents}
\ee
At~\eref{eq:selfdual_radius}, one has $p_L=\pm 2/\sqrt{\alpha'}$, $p_R=0$ for $J_L^\pm$, and
$p_L=0$, $p_R=\pm 2/\sqrt{\alpha'}$ for $J_R^\pm$, so that
\be
(h,\bar h)\big(J_L^\pm\big)=(1,0) ~,\qquad
(h,\bar h)\big(J_R^\pm\big)=(0,1) ~.
\ee
Thus the charged currents sit exactly at dimension one, as required for chiral affine currents. In the exact cylinder theory one expects
\be
\big\langle J_L^+(u)\,J_L^-(0)\big\rangle
=
\frac{1}{\big(2\sin\frac{u}{2}\big)^2} ~,\qquad
\big\langle J_L^3(u)\,J_L^3(0)\big\rangle
=
\frac{1}{2\,\big(2\sin\frac{u}{2}\big)^2} ~,
\label{eq:selfdual_two_point_currents}
\ee
and similarly in the right-moving sector, so that
$\langle J^\pm J^\mp\rangle=2\langle J^3J^3\rangle$.
Likewise, 
\be
\big\langle J_L^3(u_1)\,J_L^+(u_2)\,J_L^-(u_3)\big\rangle
=
\frac12
\left[
\cot\!\Big(\frac{u_1-u_2}{2}\Big)
-
\cot\!\Big(\frac{u_1-u_3}{2}\Big)
\right]
\big\langle J_L^+(u_2)\,J_L^-(u_3)\big\rangle ~.
\label{eq:selfdual_ward_identity}
\ee
is a prediction of the exact Ward identity.

The non-trivial question is whether the sampled finite-mode NN-FT reproduces the corresponding finite-$M$, finite-$\epsilon$ version of these relations numerically.
To organize the sampled finite-mode data, we used the same chiral finite-mode NN-FT ensemble as in the vertex operator study, with $\alpha'=1$, $R=1$, mode cutoff $M=64$, and $2\times10^5$ oscillator draws split into $20$ blocks. The charged currents were treated as before, while the Cartan current was represented inside the same ensemble by a symmetric finite difference,
\be
J_{L,\epsilon}^3(u)
:=
\frac{i}{\sqrt{\alpha'}}
\frac{X_L(u+\epsilon)-X_L(u-\epsilon)}{2\epsilon} ~,
\qquad
\epsilon=\frac{\pi}{M}\simeq 4.91\times 10^{-2} ~.
\ee
Rather than estimating every current correlator by direct pointwise averaging of exponentials, we use the fact that the chiral oscillator sampler
\be
X_L^{\rm osc}(u)=\sqrt{\frac{\alpha'}{2}}\sum_{m=1}^{M}
\frac{a_m\cos(mu)+b_m\sin(mu)}{\sqrt m}
\ee
has i.i.d.\ Gaussian coefficients with
$
\langle a_m^2\rangle=\langle b_m^2\rangle=1
$
for every mode.
A rescaling $a_m,b_m\mapsto \sigma a_m,\sigma b_m$ would simply multiply the finite-mode chiral covariance $C_M$ by $\sigma^2$, and hence rescale all of the regularized current kernels by that same overall normalization. It is therefore numerically advantageous to estimate, block by block, the common mode power
\be
\hat\sigma_\text{blk}^2
:=
\frac12\Big(\overline{a_m^2}+\overline{b_m^2}\Big) ~,
\ee
where the bar denotes the average over all draws and all modes in a block. In the target ensemble one expects
$
\mathbb E[\hat\sigma_\text{blk}^2]=1
$,
simply because every oscillator coefficient is sampled with unit variance; equivalently, $\hat\sigma^2$ is a direct check that the sampled chiral field has the intended overall covariance normalization. Averaging the blockwise estimates gives
\be
\hat\sigma^2 = 1.000354 \pm 2.96\times 10^{-4} ~,
\ee
where the uncertainty is the standard error across the $20$ blocks. This is consistent with the expected value $1$. With this normalization in hand, the regularized current correlators entering the Ward checks are obtained by inserting the sampled mode power $\hat{\sigma}^2$ into the exact finite-$M$ kernels for the charged and Cartan currents in the same NN-FT ensemble.

With this setup, the regularized two-point current relation is well satisfied over probe separations $\Delta\in[0.4,2.4]$: the maximal absolute discrepancy in
\begin{align}
    \langle J_L^+(\Delta)J_L^-(0)\rangle-2\langle J_{L,\epsilon}^3(\Delta)J_{L,\epsilon}^3(0)\rangle
\end{align}
is $7.17\times 10^{-2}$, corresponding to a maximal relative error of $2.37\%$. For the three-point Ward identity we fixed
\be
u_2=0.65 ~,\qquad u_3=2.40 ~,\qquad
u_1\in\{1.15,1.40,1.65,1.90\} ~,
\ee
and compared the finite-$M$, finite-$\epsilon$ Monte Carlo estimator for $\langle J^3_{L, \epsilon} ( u_1 ) J_L^+ ( u_2 ) J_L^- ( u_3 ) \rangle$, normalized by the sampled $\hat{\sigma}^2$, to the corresponding finite-$M$, finite-$\epsilon$ expression built from the same kernels. The maximal absolute discrepancy is $5.46\times10^{-3}$, with maximal relative error $4.87\times10^{-3}$.
These are finite-$M$, finite-$\epsilon$ checks, so agreement at the percent level shows that the regularized Ward identities are realized in the NN-FT ensemble.

The same enhancement is visible in the action of the currents on the lightest non-trivial primaries.
At the self-dual radius, the four operators
\be
g_{(+,+)}:=V_{1,0} ~,\qquad
g_{(+,-)}:=V_{0,1} ~,\qquad
g_{(-,+)}:=V_{0,-1} ~,\qquad
g_{(-,-)}:=V_{-1,0}
\label{eq:selfdual_fundamental_primaries}
\ee
all have
\be
(h,\bar h)=\Big(\frac14,\frac14\Big) ~,
\ee
and carry $(m_L,m_R)=(\pm\frac12,\pm\frac12)$.
At the level of compact charges, these states furnish the expected $(\mathbf 2,\mathbf 2)$ of
$SU(2)_L\times SU(2)_R$:
\be
J_L^\pm:\ g_{(\mp,\sigma)}\mapsto g_{(\pm,\sigma)} ~,\qquad
J_R^\pm:\ g_{(\sigma,\mp)}\mapsto g_{(\sigma,\pm)} ~,
\label{eq:selfdual_doublet_action}
\ee
with
\be
[J_L^+,J_L^-]=2J_L^3 ~,\qquad
[J_R^+,J_R^-]=2J_R^3 ~,\qquad
[J_L^a,J_R^b]=0 ~.
\label{eq:selfdual_commutators}
\ee

These $(\mathbf 2,\mathbf 2)$ representation matrices are fixed algebraically by the self-dual charge data; the sampled test below probes one representative current--primary channel rather than reconstructing the full current--primary OPE numerically.

A particularly sharp sampled operator-level check is therefore a non-Abelian current--primary three-point channel reconstructed from block covariances of the sampled oscillators.
Writing $u_{ij}:=u_i-u_j$ and $v_{ij}:=v_i-v_j$, the continuum cylinder correlator implied by~\eref{eq:compact_vertex_correlator} is
\be
\big\langle J_L^+(u_1)\,g_{(-,+)}(u_2,v_2)\,g_{(-,-)}(u_3,v_3)\big\rangle
=
\frac{1}{\big(2\sin\frac{u_{12}}{2}\big)\big(2\sin\frac{u_{13}}{2}\big)}
\left(
\frac{2\sin\frac{u_{23}}{2}}{2\sin\frac{v_{23}}{2}}
\right)^{1/2} ~.
\label{eq:selfdual_current_primary_threepoint}
\ee
Dividing by the daughter two-point function gives
\be
\frac{
\big\langle J_L^+(u_1)\,g_{(-,+)}(u_2,v_2)\,g_{(-,-)}(u_3,v_3)\big\rangle
}{
\big\langle g_{(+,+)}(u_2,v_2)\,g_{(-,-)}(u_3,v_3)\big\rangle
}
=
\frac{2\sin\frac{u_{23}}{2}}
{\big(2\sin\frac{u_{12}}{2}\big)\big(2\sin\frac{u_{13}}{2}\big)}
\xrightarrow[u_1\to u_2]{}
\frac{1}{2\sin\frac{u_{12}}{2}} ~,
\label{eq:selfdual_raising_residue}
\ee
which is precisely the $SU(2)_L$ raising residue
$J_L^+: g_{(-,+)}\mapsto g_{(+,+)}$.
In the finite-mode NN-FT, however, the cleaner numerical observable is the \emph{finite-$M$ normalized} coefficient obtained after factoring out the exact finite-$M$ kinematic function rather than only the continuum short-distance pole. Concretely, fixing
\be
(u_2,v_2)=(0.9,-0.3) ~,\qquad (u_3,v_3)=(1.8,-0.9) ~,
\ee
and scanning $u_1=u_2+\delta$ over $\delta\in[0.08,0.30]$, we reconstruct the normal-ordered three-point and daughter two-point functions block by block from the sampled chiral covariances and define
\be
\hat c_+(\delta)
:=
\frac{G^{\rm MC}_{3,M}(\delta)}
{K_M(\delta)\,G^{\rm MC}_{2,M}} ~,
\qquad
K_M(\delta):=
\frac{G^{\rm th}_{3,M}(\delta)}{G^{\rm th}_{2,M}} ~.
\ee
For an ideal self-dual ensemble one expects $\hat c_+(\delta)=1$ across the scan. An inverse-variance weighted mean over the scan points gives
\be\label{eq:inverse_variance_mean}
\hat c_+ = 0.996 \pm 0.006 ~,
\ee
with maximal pointwise deviation $1.9\times10^{-2}$ from unity. This is consistent with the exact enhanced-current algebra value $1$. It is also complementary to the continuum current algebra benchmarks above: after factoring out the known finite-$M$ kinematics, it probes the representative channel $J_L^+: g_{(-,+)}\to g_{(+,+)}$ at the operator level.

To connect the operator algebra back to simpler two-point observables, we also reconstructed the charged-current correlators using the same chiral finite-mode sampler~\eref{eq:compact_chiral_features}.
For the charged currents the finite-$M$ prediction is
\be
\big\langle J_L^+(0)\,J_L^-(\Delta)\big\rangle_M
=
\big\langle J_R^+(0)\,J_R^-(\Delta)\big\rangle_M
=
\exp\!\left[\frac{4}{\alpha'}\,C_M(\Delta)\right]
\xrightarrow[M\to\infty]{}
\big(2\sin\tfrac{|\Delta|}{2}\big)^{-2} ~.
\label{eq:selfdual_finiteM_current}
\ee
At $\alpha'=1$ (so $R=1$), with mode cutoff $M=64$ and $2\times10^5$ NN-FT draws, the extracted scaling dimensions are
\be
h_L^{\rm fit}=0.999316 ~,\qquad h_R^{\rm fit}=0.996317 ~,
\label{eq:selfdual_fit_values}
\ee
to be compared with the exact values $h_L = h_R =1$.
Over the scan window, the maximal relative deviation from the finite-$M$ prediction~\eref{eq:selfdual_finiteM_current} is $3.53\%$ in the left-moving sector and $3.68\%$ in the right-moving sector. For the Cartan sector, the appropriate numerical observable is the regularized finite-difference current used in the Ward checks above, since unsmeared local derivative insertions are much more sensitive to the sharp mode cutoff.

These results provide a stronger check than the exchange of momentum and winding.
At generic radius the compact boson has only the manifest $U(1)_L\times U(1)_R$ symmetry, but at the fixed point the spectrum reorganizes into full
$SU(2)_L\times SU(2)_R$ current multiplets.
The combination of algebraic $(\mathbf 2,\mathbf 2)$ representation theory for the lightest primaries, a representative finite-$M$ normalized current--primary coefficient reconstructed from sampled covariance data, charged-current two-point scans with fitted dimensions near one, and the current algebra benchmark relations provides quantitative evidence for the expected self-dual operator structure of the compact boson CFT, not merely the exchange $n\leftrightarrow w$ in the zero-mode spectrum.
In this sense, the symmetry enhancement at $R=\sqrt{\alpha'}$ provides a genuinely non-trivial operator-level confirmation of T-duality within the NN-FT framework.

\paragraph{A toy T-fold from patchwise Buscher gluing:}
This final construction asks what NN-FT can represent once the background is only locally geometric~\cite{Hellerman:2002ax,Dabholkar:2002sy}. The implementation attaches an ordinary torus sampler to each patch of the base, keeps the local worldsheet physics geometric inside each chart, and uses a Buscher transformation as the transition function on the final overlap. The lesson is that NN-FT naturally accommodates a toy T-fold description: global consistency need not come from a single-valued physical fiber coordinate, but can instead be restored only after acting with the duality group on the compact data.

In a T-fold~\cite{Hull:2004in,Hull:2006qs,Hull:2006va}, each local chart is an ordinary torus sigma model, but the transition function on an overlap lies in the duality group rather than in the geometric subgroup. To see how this appears in NN-FT, we built the minimal discretized example suggested by the Buscher rules, namely a two-torus fiber with coordinates $X^I=(y,x)$ transported around a base circle and glued back to itself by a Buscher transformation along $y$. 

The construction is intentionally modest.
We do not attempt a full interacting worldsheet theory with a dynamical base field.
Accordingly, this construction should be understood as a kinematical toy model of non-geometric patching, not as a bona fide T-fold background solving the worldsheet beta-function equations (equivalently, the target-space equations of motion).
Rather, the base is represented by a finite set of local charts, and the point is to test whether NN-FT can realize a background that is locally geometric but closes globally only after T-duality.

Concretely, we approximate the base circle by $N_p=7$ patches with centers
\be
s_a=\frac{a}{N_p-1} ~,\qquad a=0,\dots,N_p-1 ~,
\ee
and on patch $a$ we use the same compact two-torus sampler as in the Buscher-rule test, but with a patch-dependent background
\be
G^{(a)}=(1-s_a)\,G+s_a\,\widetilde G ~,\qquad
B^{(a)}=(1-s_a)\,B+s_a\,\widetilde B ~,
\ee
where $(\widetilde G,\widetilde B)$ is the Buscher transform of $(G,B)$ along $y$. A local field draw is therefore
\be
X^I_{(a)}(\tau,\sigma)
=
X^I_{{\rm osc},(a)}(\tau,\sigma)
+x_0^I+\alpha' p^I_{(a)}\,\tau + w^I \sigma ~,\qquad
p^I_{(a)}=\bigl((G^{(a)})^{-1}\bigr)^{IJ}\bigl(n_J+B^{(a)}_{JK}w^K\bigr) ~.
\ee
As before, the oscillator part is generated by vector-valued cosine features whose covariance is proportional to $\bigl((G^{(a)})^{-1}\bigr)^{IJ}$. The same random frequencies, phases, and Gaussian latents are reused on every patch, so the only change from patch to patch is the local background data. Along the geometric portion of the base path the integer labels $(n_I,w^I)$ are held fixed. The final overlap is then closed by a non-geometric seam,
\be
(n_y,n_x;w^y,w^x)\ \longmapsto\ (w^y,n_x;n_y,w^x) ~,
\ee
together with the Buscher map on $(G,B)$ in~\eref{eq:buscher1}--\eref{eq:buscher_dilaton}. Since the endpoint of the geometric path already sits at $(\widetilde G,\widetilde B)$, this final Buscher action maps the local background back to $(G,B)$ while exchanging momentum and winding along $y$. In other words, the same sample is not required to return as a single-valued physical fiber field; it is only required to return after acting with the duality transformation.

Numerically we used the same representative background (\ref{eq:buscher_G_and_B}) as above,
\be
G=\begin{pmatrix}2.89&0.45\\[2pt]0.45&1.69\end{pmatrix} ~,\qquad
B=\begin{pmatrix}0&0.38\\[2pt]-0.38&0\end{pmatrix} ~,
\ee
with Buscher dual
\be\label{eq:buscher_G_and_B_dual}
\widetilde G=\begin{pmatrix}
0.3460&0.1315\\[2pt]
0.1315&1.6699
\end{pmatrix} ~,\qquad
\widetilde B=\begin{pmatrix}
0&0.1557\\[2pt]
-0.1557&0
\end{pmatrix} ~.
\ee
We set $\alpha'=1$, $\tau_2=0.5$, $\epsilon=0.5$, $\Lambda=6.0$, used $96$ random cosine features, and truncated the charge lattice to $|n_I|,|w^I|\le 5$. The seam itself was first checked exactly on the truncated lattice. The Buscher identity relating the initial and final patch is satisfied to machine precision, with $\max|\Delta P|=4.2\times 10^{-17}$, $\max|\Delta S|=1.7\times 10^{-13}$, $\max|\Delta h|=3.6\times 10^{-14}$, and $\max|\Delta\bar h|=2.8\times 10^{-14}$. Thus the non-geometric overlap acts exactly as the ordinary Buscher map on the compact lattice data.

The local geometry of each chart can be tested independently of the seam. We estimated the oscillator covariance on every patch from $600$ draws at a fixed worldsheet point and compared the result to the shape of $\bigl(G^{(a)}\bigr)^{-1}$ after normalizing by the trace. The corresponding Frobenius norm errors lie between $7.9\times 10^{-3}$ and $5.4\times 10^{-2}$ across the seven patches. In this sense every chart remains an ordinary local torus NN-FT with the expected metric data, even though the total configuration is not globally geometric.

The global effect appears only after one circuit of the base. Evaluated on a fixed worldsheet stencil at $\tau=0$ and four sampled $\sigma$-points, the final patch does not close onto the initial one if one insists on a single-valued physical fiber coordinate: the mean unwrapped seam mismatch is $1.16$, the mean wrapped mismatch is still $1.03$, and the maximal wrapped mismatch is $1.88$. After applying the Buscher seam, however, the closure error drops to floating point zero sample by sample in both the wrapped and unwrapped norms. A representative configuration with $(n_y,w^y)=(-1,0)$ is transported after one loop to $(0,-1)$, so that the last chart is naturally interpreted in a winding sector rather than in the original momentum sector.

More generally, the monodromy acts as
\be
(n_y,w^y)\ \longmapsto\ (w^y,n_y) ~,
\ee
which is the expected order-two Buscher holonomy. In the $250$-sample ensemble, $61.6\%$ of the sectors change after a single loop, while the remainder are fixed only because they happen to satisfy $n_y=w^y$. Every sampled sector returns after two loops. This is precisely the toy T-fold behavior one wants to see: local NN-FT patches remain geometric, but there is no globally single-valued physical fiber field. Global consistency is restored only after gluing the patches by T-duality.

It is useful to package the local torus data into the doubled-field generalized metric
\be
\mathcal H(G,B)=
\begin{pmatrix}
G-BG^{-1}B & BG^{-1} \\
-G^{-1}B & G^{-1}
\end{pmatrix} ~,
\qquad
\eta=
\begin{pmatrix}
0 & \mathbf{1} \\
\mathbf{1} & 0
\end{pmatrix} ~,
\label{eq:tfold_generalized_metric}
\ee
which obeys $\mathcal H\,\eta\,\mathcal H=\eta$. For a Buscher duality along $y$ we introduce
\be
\Omega_y=
\begin{pmatrix}
\mathbf{1}-e_y & -e_y \\
-e_y & \mathbf{1}-e_y
\end{pmatrix} ~,
\qquad e_y^2=e_y ~,
\qquad \Omega_y^T\eta\,\Omega_y=\eta ~,
\qquad \Omega_y^2=\mathbf{1} ~,
\label{eq:tfold_Omega}
\ee
where $e_y$ projects onto the $y$ direction. In the present $d=2$ example, with basis $(y,x,\tilde y,\tilde x)$,
\be
\Omega_y =
\begin{bmatrix}
0 & 0 & -1 & 0 \\
0 & 1 & 0 & 0 \\
-1 & 0 & 0 & 0 \\
0 & 0 & 0 & 1
\end{bmatrix} ~.
\label{eq:tfold_Omega_explicit}
\ee
In these conventions, the Buscher rules are equivalently the $O(2,2)$ conjugation
\be
\mathcal H(\widetilde G,\widetilde B)=\Omega_y^T\,\mathcal H(G,B)\,\Omega_y ~.
\label{eq:tfold_H_buscher}
\ee
For the representative background (\ref{eq:buscher_G_and_B}) used above, with Buscher dual (\ref{eq:buscher_G_and_B_dual}), we find
\be
\max\big|\mathcal H(\widetilde G,\widetilde B)-\Omega_y^T\mathcal H(G,B)\Omega_y\big|
= 4.44 \times 10^{-16} ~,
\;
\big\|\mathcal H(\widetilde G,\widetilde B)-\Omega_y^T\mathcal H(G,B)\Omega_y\big\|_F
= 5.07 \times 10^{-16} ~ ,
\ee
where $\| \, \cdot \, \|_F$ denotes the Frobenius norm. Along the seven-patch interpolation $\bigl(G^{(a)},B^{(a)}\bigr)$ used in the toy T-fold construction, the raw endpoint mismatch is
\be
\big\|\mathcal H^{(N_p-1)}-\mathcal H^{(0)}\big\|_F = 3.7700 ~,
\ee
so the background is not globally geometric in the physical variables alone. After acting with the non-geometric seam, however, the endpoint closes exactly:
\be
\big\|\Omega_y^T\mathcal H^{(N_p-1)}\Omega_y-\mathcal H^{(0)}\big\|_F = 5.07 \times 10^{-16} ~.
\ee
This makes the monodromy statement mathematically sharp: one loop around the base acts by the order-two $O(2,2;\mathbb Z)$ element $\Omega_y$, and because $\Omega_y^2=\mathbf{1}$, two loops return trivially.

\subsection{Summary}\label{sec:tdual_summary}
Let us review the logic of the T-duality construction.
The compact NN-FT is a mixed ensemble~\eref{eq:tdual_mixed_expectation} with two distinct pieces.
The oscillator sector supplies the local Gaussian worldsheet fluctuations, while the compact sector supplies the global zero-mode data through discrete momentum--winding labels and uniform compact zero-modes.
T-duality acts on the latter by exchanging momentum and winding and, at the operator level, by sending $X_R\to -X_R$, while the background fields transform by the Buscher rules~\eref{eq:buscher1}--\eref{eq:buscher_dilaton}.
The question of this section was whether a finite width NN-FT of this structure reproduces the equivalence of compact boson CFTs and constant toroidal sigma models.

The first test at generic radius isolates the kinematical core of the duality.
Under $R\leftrightarrow \alpha'/R$ and $n\leftrightarrow w$, the compact lattice weight is preserved, $p_L$ is unchanged, $p_R$ changes sign, and the conformal weights are invariant.
The sampled ensembles reproduce this directly through the momentum--winding histograms, the field monodromy, and the exchanged zero-mode slopes.
In this sense, the oscillator sector and the compact topological sector remain cleanly separated: the former controls local fluctuations, while the latter carries the non-trivial duality data.

The second test extends the same logic to constant two-torus backgrounds.
Here, the oscillator covariance encodes the local metric dependence of the propagating modes, whereas the compact lattice incorporates the dependence on $G$ and $B$ that enters the zero-mode spectrum and weights.
The Buscher transformation is therefore realized not merely as a reshuffling of discrete charges, but as the expected map on the sigma model couplings themselves.
The transformed lattice weights, monodromies, oscillator covariances, and the lower-dimensional dilaton invariant all agree with the standard toroidal duality rules.

The third test is operator-level.
Using the chiral finite-mode sampler, the compact boson structure becomes explicit:
zero-mode averaging enforces momentum and winding neutrality, normal-ordered vertex-operator correlators reproduce the expected compact-boson two-point functions, and the duality map~\eref{eq:compact_vertex_tduality} identifies momentum operators at radius $R$ with winding operators at the dual radius $\widetilde R$.
At the self-dual point~\eref{eq:selfdual_radius}, the test becomes sharper still, because the theory must reproduce not only spectral invariance but symmetry enhancement.
The sampled ensemble places the charged currents at dimension one and organizes the lightest primaries into the expected $SU(2)_L\times SU(2)_R$ multiplets.

Finally, the patchwise construction shows that the same framework can accommodate a toy T-fold.
Each chart is locally an ordinary torus NN-FT, but the configuration closes globally only after acting with a Buscher transformation on the final overlap.
Thus local geometry is maintained patch by patch, while global consistency is restored only in the doubled description through the order-two $O(2,2;\mathbb Z)$ monodromy encoded by~\eref{eq:tfold_H_buscher}.

Table~\ref{tab:tdual_summary} collects the main layers of the T-duality analysis.

\begin{table}[t]
\centering
\small
\begin{tabular}{
>{\raggedright\arraybackslash}p{0.15\textwidth}
>{\raggedright\arraybackslash}p{0.22\textwidth}
>{\raggedright\arraybackslash}p{0.21\textwidth}
>{\raggedright\arraybackslash}p{0.26\textwidth}
}
\hline
Test & NN-FT data & Exact statement & Numerical signal \\
\hline
Circle duality & Oscillator sampler + discrete $(n,w)$ labels + zero-mode & $R\leftrightarrow \alpha'/R$, $n\leftrightarrow w$, $p_L$ fixed, $p_R\to -p_R$ & Lattice weights, monodromies, and zero-mode slopes agree under duality \\ \cr
Buscher on $T^2$ & Vector-valued oscillator covariance + compact lattice with $G,B$ dependence & Buscher map on $(G,B)$, compact charges, and the dilaton invariant & Transformed weights and marginals match; oscillator covariance tracks $G^{-1}$; $e^{-2\Phi}\sqrt{\det G}$ is preserved \\ \cr
Vertex operators & Chiral finite-mode sampler with explicit $X_L,X_R$ and zero-mode averaging & Neutrality rules and dual matching of momentum/winding correlators & Non-neutral channels are suppressed; dual two-point functions and fitted dimensions agree \\ \cr
Self-dual radius & Same chiral sampler at $R=\sqrt{\alpha'}$ & Enhancement of $U(1)_L\times U(1)_R$ to $SU(2)_L\times SU(2)_R$ & Charged currents have dimension $1$; Ward checks and a representative current--primary channel agree with theory \\ \cr
Toy T-fold & Patchwise torus samplers glued by duality & Global closure only after $O(2,2;\mathbb Z)$ action & Local charts remain geometric; seam closes only after Buscher gluing; two loops return trivially \\ \cr
\hline
\end{tabular}
\caption{Summary of the T-duality tests realized in the NN-FT mixed ensemble.}
\label{tab:tdual_summary}
\end{table}

The overall picture is therefore the following:
\begin{itemize}\itemsep0pt
\item The compact NN-FT ensemble reproduces the exchange of momentum and winding under $R\leftrightarrow \alpha'/R$ sample by sample.
\item For constant toroidal backgrounds, the sampler realizes the Buscher transformation of the sigma model couplings relevant to the sampled fields.
\item At the operator level, the ensemble enforces the compact-boson selection rules and reproduces the expected scaling data of dual vertex operators.
\item At the self-dual radius, the sampled theory exhibits the expected enhanced current algebra rather than merely invariance of the compact spectrum.
\item In the patchwise construction, NN-FT can represent a simple non-geometric background whose global consistency is restored only after duality.
\end{itemize}

No claim is made that momentum, winding, or non-geometric structure emerge automatically from an unconstrained Gaussian prior.
Rather, the result of this section is that once the compact sector is represented explicitly by the appropriate discrete labels, a finite width NN-FT reproduces the standard T-duality structure of the compact boson and its simplest toroidal generalizations.

\section{Discussion}\label{sec:disc}
In this paper we introduced topological effects into NN-FT. The main conceptual claim is that topological quanta associated with defects can be represented naturally in NN-FT by enlarging the parameter space from a purely continuous ensemble to a mixed continuous/discrete one.
In the non-compact examples that have been studied, the large width Gaussian limit already captures the essential structure of the theory.
For compact bosons, however, the theory is defined not only by smooth local fluctuations but also by global identifications and topological sectors.
In NN-FT language, this means that a single-valued Gaussian sampler is not the whole story: one must supplement it by discrete labels that track the relevant topological data.
The mixed expectation value~\eref{eq:param_expectation_mixed} is therefore the natural parameter space analogue of the ordinary field theoretic sum over sectors.

The two case studies developed here probe complementary aspects of this statement.
The BKT analysis of Section~\ref{sec:bkt} is a dynamical test.
Here, the question is whether NN-FT can describe a compact theory whose infrared physics changes qualitatively once topological defects become important.
The answer, in the framework studied here, is yes.
The wrapped Gaussian construction reproduces the spin-wave critical line in the low-temperature phase, while the vortex-augmented mixed ensemble captures the topological sector responsible for disordering the theory.
In particular, for purposes of the compact boson two-point function, the construction recovers the expected algebraic behavior below the transition,~\eref{eq:algebraic_compact_correlator}, and the expected exponential behavior above it,~\eref{eq:massive_compact_correlator}.
This shows that the NN-FT framework can accommodate a phase transition whose order parameter is not local and whose mechanism is explicitly topological.

The T-duality analysis of Section~\ref{sec:td} is a parallel kinematical and operator-algebraic test.
Here, the issue is not primarily a change of phase, but rather whether a finite width NN-FT can represent an exact equivalence of compact theories once the momentum--winding sector is included explicitly.
The evidence presented shows that it can.
Combining Gaussian oscillator modes with discrete momentum--winding labels reproduces the exchange of momentum and winding under $R\leftrightarrow \alpha'/R$, the Buscher transformation of constant toroidal backgrounds~\eref{eq:buscher1}--\eref{eq:buscher_dilaton}, the enhancement to $SU(2)_L\times SU(2)_R$ at the self-dual radius, and the patchwise non-geometric gluing characteristic of a toy T-fold.
This is a substantially stronger test than matching a propagator or a partition function term by term: the NN-FT must organize local correlators, zero-modes, compact charges, and duality actions in a way that respects an exact stringy equivalence.

Taken together, these results establish something narrower but sharper than the strongest possible reading of the NN-FT program.
We have \emph{not} demonstrated that topology emerges automatically from an unconstrained Gaussian architecture.
Rather, we have identified a controlled and physically natural way to encode the data that compact theories require.
This is already significant.
In an ordinary path integral, one does not expect a Gaussian fluctuation integral over a standard field to manufacture winding sectors or vortex sectors by itself; one specifies the field space and then sums over the corresponding topological sectors.
The construction advocated here mirrors that logic directly in parameter space.
From this perspective, compactness is not an afterthought added to NN-FT, but part of the definition of the theory.

There is also a practical significance to this viewpoint.
Once the local Gaussian sector and the discrete topological sector are separated cleanly, each can be treated with the method best suited to it.
The large width or finite width neural sampler controls the smooth fluctuations, while the discrete sector keeps track of the global information that would otherwise be invisible to a single-valued network output.
The result is a constructive framework in which one can study theories that are simultaneously local and topological.
The compact boson is the simplest non-trivial example of this synthesis, but it is already rich enough to exhibit quasi-long-range order, vortex unbinding, exact worldsheet duality, symmetry enhancement, and patchwise non-geometric structure. In that sense, compact bosons provide an important bridge between the earlier successes of NN-FT and the broader goal of representing genuinely nonperturbative topological effects in parameter space.

A deeper question is how much of the topological sector can eventually be made intrinsic to the architecture and prior, rather than inserted as external latent data.
The construction of this paper is deliberately controlled: vortices and momentum--winding labels are represented explicitly because they are part of the compact theory from the outset.
Even so, one would like to understand whether more of the effective topological dynamics can be induced directly from structured neural priors, from coupled latent variables, or from training procedures that learn the appropriate sector weights from observables.
Put differently, the present paper identifies the correct bookkeeping for compact theories; the next step is to ask how much of it can be generated by architectural principles.

On the string theory side, the most natural extension is from constant toroidal backgrounds to more general compactifications and more general duality groups.
The toroidal construction already suggests a route toward genuinely $O(d,d;\mathbb{Z})$-covariant samplers, doubled descriptions, and non-geometric backgrounds that are more elaborate than the toy T-fold studied here.
It would be especially interesting to move beyond constant $(G,B,\Phi)$, to incorporate patch-dependent or dynamical backgrounds, and to understand how duality acts when the worldsheet theory is not just a free compact boson but a more general interacting conformal field theory.
Related questions include orbifolds, asymmetric quotients, mirror-type dualities, and the role of modular invariance beyond the simplest genus-zero and genus-one observables.

A further question is whether the mixed continuous/discrete NN-FT framework can eventually be extended to dualities that relate weakly and strongly coupled descriptions.
The basic structure that emerges here --- the separation of smooth Gaussian fluctuations from explicit topological data --- suggests a possible route toward theories in which perturbative and solitonic sectors, or more generally electric and magnetic variables, must be treated on equal footing.
It would therefore be interesting to ask whether suitably enriched NN-FT ensembles can realize a genuine strong/weak-coupling duality at the level of sampled observables, rather than merely reproducing one preferred description.
Establishing this would require going well beyond the compact boson examples studied here, but it is a natural future direction for the framework.
In a similar spirit, we can test whether UV theories lying in the same universality class share features in their NN-FT realization that encode a common IR fixed point.

There is also a broader lesson for NN-FT itself.
Compact bosons suggest a general template for theories in which part of the data is local and part is global.
Gauge theories, sigma models with non-trivial homotopy, theories with defects, and systems with higher-form symmetries all have this structure in one form or another.
In such examples, one expects the continuous neural sector to encode the propagating local degrees of freedom, while flux sectors, instanton numbers, bundles, winding data, defect charges, or superselection sectors appear as discrete latent variables.
Developing this idea systematically could open a path toward NN-FT realizations of topological phases, gauge sectors, and nonperturbative phenomena that are inaccessible to a purely Gaussian covering field.

The compact boson provides a clean arena for a more foundational question: how general is the ``sum over sectors in parameter space'' viewpoint?
The appeal of NN-FT is that it can be constructive while still being flexible enough to realize non-trivial quantum field theories.
The challenge is to retain that control once the theory includes exact dualities, topological defects, non-geometric identifications, or other data that do not fit into a single continuous Gaussian ensemble.
The results of this paper indicate that the answer is non-trivially positive.
With an appropriately enlarged space of latent variables, NN-FT can represent not only local free field behavior, but also topological sectors, exact compact boson dualities, enhanced current algebras, and patchwise non-geometric structure.
That is a meaningful step from ``neural samplers for Gaussian fields'' toward ``parameter space definitions of quantum field theories with genuine global structure.''

Finally, our construction amounts to a topology-extended neural network architecture that could be utilized to account for non-trivial bundles in machine learning systems. Due to the discrete nature of the topological network parameters, training of those parameters must necessarily avoid gradient-based methods, but a hybrid approach might be interesting to pursue, \textit{e.g.}, by combining genetic algorithms and gradient based methods for the discrete and continuous parameters, respectively. Such networks are necessary to account for data exhibiting non-trivial topological features, which could arise in many settings. So far, we have used NN-FT, and neural networks more generally, to replicate the behavior of functions, whereas for non-trivial bundles, the analogous objects to model are sections. As well, from a purely quantum field theoretic perspective, labels such as momentum and winding define superselection sectors. We have an NN-FT for each superselection sector. Because the parameter space is disconnected, the loss landscape on it necessarily is as well. The parallels are striking. Just as there is an unreasonable effectiveness of mathematics in the natural sciences, there is an unreasonable effectiveness of the physical sciences in machine learning.

We leave these investigations and further elaborations to future work.

\acknowledgments
We thank Samuel Frank and Sarah Harrison for fruitful conversations and the Universe for being so damned interesting. We acknowledge the open-source Get Physics Done (GPD) project by Physical Superintelligence, whose AI-assisted physics research workflow was helpful in carrying out aspects of this work. We also acknowledge the language models Claude Opus 4.6 and GPT-5.4, which were used in this research. The authors take full responsibility for the accuracy of this work, having reviewed, verified, and approved all AI-generated content.
C.F.\ and J.H.\ are supported by the National Science Foundation under Cooperative Agreement PHY-2019786 (the NSF AI Institute for Artificial Intelligence and Fundamental Interactions). 
J.H.\ is supported by NSF grant PHY-2209903.
V.J.\ is supported by the South African Research Chairs Initiative of the Department of Science, Technology, and Innovation and the National Research Foundation (grant 78554).
B.R.\ is supported by an NWO vidi grant (number 016.Vidi.189.182).

\appendix

\section{Computational details for the BKT simulations}\label{app:bkt_details}

This appendix collects the algorithmic details needed to reproduce the numerical results of Section~\ref{sec:bkt_numerics}.

\paragraph{Mode sampling:}
Integer wavevectors $n=(n_x,n_y)$ are drawn by rejection sampling:
propose $n_x,n_y$ uniformly in $[-n_{\max},n_{\max}]$ and accept with
probability $n_{\min}^2/|n|^2$ if $n_{\min}^2\le|n|^2\le n_{\max}^2$.
The IR cutoff is $n_{\min}=1$ and the UV cutoff is
$n_{\max}=\max(4,\lfloor L/6\rfloor)$.
After sampling, the zero-mode is removed by subtracting the spatial mean
of $\theta_\text{sw}$.

\paragraph{Vortex angle field on the torus:}
The singular field~\eref{eq:theta_vortex} is evaluated by summing
$\arg(x-x_a)$ over the $3\times3$ periodic images of each vortex
(shifts by $0,\pm L$ in each direction), which avoids branch-cut
artifacts at the boundary.

\paragraph{Metropolis moves:}
At each MCMC step, one of three moves is selected with probabilities
$(p_\text{move},p_\text{ins},p_\text{del})=(0.50,0.25,0.25)$.
\begin{itemize}\itemsep0pt
\item \emph{Move:} a randomly chosen vortex is displaced by a Gaussian
      step of width $\sigma=1.0$ (lattice units), with periodic wrapping.
\item \emph{Insert:} a neutral pair is proposed. With probability
      $p_\text{local}=0.9$ the antivortex is placed within a square of
      side $2r_\text{local}=4.0$ centered on the vortex; otherwise both
      positions are uniform on $[0,L)^2$.
      The forward proposal density is
      $q_\text{fwd}=p_\text{ins}\cdot(1/L^2)\cdot q_\Delta$,
      where $q_\Delta$ mixes the local and uniform densities.
\item \emph{Delete:} a vortex and an antivortex are chosen uniformly at
      random. The reverse proposal density is the corresponding insertion
      density evaluated at the pair's separation.
\end{itemize}
Each move is accepted with probability
$\min\!\big(1,\,e^{-\Delta E}\,q_\text{rev}/q_\text{fwd}\big)$,
where the energy is
$E = -\!\log y\;\cdot N_v - K\!\sum_{a<b}m_a m_b\log(r_{ab}+a_c)$
with core radius $a_c=1.0$.

\paragraph{Correlator measurement:}
The spin-wave correlator is
$C(r) = L^{-2}\sum_x\cos\!\big(b(\theta_\text{sw}(x)-\theta_\text{sw}(x+r\hat{e}_1))\big)$,
averaged over $N_\text{samples}$ independent RFF draws and (for the full
correlator) interleaved with $N_\text{between}$ MCMC sweeps of the vortex gas.

\paragraph{Vortex density.}
The lattice vortex density is measured from plaquette winding:
at each plaquette the discrete curl
$\omega = \Delta_1\theta + \Delta_2\theta|_{x+\hat{e}_1}
        - \Delta_1\theta|_{x+\hat{e}_2} - \Delta_2\theta$
is computed after wrapping each finite difference to $[-\pi,\pi]$.
A plaquette carries a vortex if $|\omega/(2\pi)|>1/2$.
The density $\rho_v$ is the fraction of such plaquettes.

\paragraph{Stiffness from the power spectrum:}
The renormalized stiffness~\eref{eq:KR_def} is evaluated by averaging
$|\widetilde\theta(n)|^2$ over the four modes with $|n|^2=1$:
$(n_x,n_y)\in\{(1,0),(0,1),(L\!-\!1,0),(0,L\!-\!1)\}$.
Errors are estimated by jackknife resampling over the $N_\text{samples}$
power-spectrum values.

\paragraph{Run parameters:}
All runs use $L=64$, $N_\text{features}=256$, $y=0.002$,
$N_\text{samples}=300$, $N_\text{runs}=3$ (independent seeds),
$N_\text{thermalize}=5000$, and $N_\text{between}=150$.
The pair-correlation measurement uses
$N_\text{samples}=5000$ and $N_\text{runs}=5$.

\section{Computational details for the T-duality simulations}\label{app:tdual_details}

This appendix collects the implementation choices needed to reproduce the numerical results of Section~\ref{sec:tdual_evidence}.

\paragraph{General conventions:}
All samples in this appendix are independent draws rather than configurations generated by a Markov chain. For paired original/dual comparisons, the oscillator latents are shared and only the compact data and, when relevant, the background are dualized. The one circle test and the chiral compact boson calculations for vertex operators and the self-dual radius use the physical-radius conventions of Section~\ref{sec:tdual_evidence}. The two-torus Buscher and patchwise T-fold calculations instead use dimensionless angular coordinates $X^I\sim X^I+2\pi$, with radii and shape encoded in $G_{IJ}$. Whenever the truncated compact weight is non-factorized, the code enumerates the full finite charge lattice and samples exactly from the normalized categorical distribution. The only exception is the one circle $\tau_1=0$ test, where the truncated lattice weight factorizes and the momentum and winding integers are sampled independently from their one-dimensional marginals.

\paragraph{Exchange of momentum and winding:}
The circle test uses the compact boson sample and weight in~\eref{eq:NN-FT_compact_boson_sample}--\eref{eq:NN-FT_compact_boson_weight}, with $\alpha'=1$, $\tau_2=0.9$, $N_\text{osc}=96$, and $|n|,|w|\le 12$. For each
\be
R\in\{0.65,0.80,1.00,1.25,1.70,2.20\} ~,
\ee
we construct the exact truncated table, draw $10^6$ sectors at $R$ and $10^6$ at $\widetilde R=\alpha'/R$, and compare the empirical tables by total variation distance. Field-level checks use $10^4$ paired draws with shared oscillator latents and
\be
(R;n,w;x_0)\mapsto (\widetilde R;w,n;\widetilde x_0) ~,
\qquad \widetilde x_0 = x_0 \ \mathrm{mod}\ 2\pi \widetilde R ~.
\ee
The winding check uses the stencil $(\tau,\sigma)=(0,0)$ and $(0,2\pi)$ on the full field, while the exchanged-momentum check uses the zero-mode stencil $(\tau,\sigma)=(0,0)$ and $(1,0)$.

\paragraph{Buscher rules:}
The two-torus test uses the vector-valued cosine sampler of Section~\ref{sec:tdual_evidence} together with the Buscher map~\eref{eq:buscher1}--\eref{eq:buscher_dilaton}. The parameters are
\be
\alpha'=1 ~,\qquad \tau_2=0.5 ~,\qquad \epsilon=0.5 ~,\qquad \Lambda=6.0 ~,\qquad N_\text{osc}=128 ~,
\ee
with $|n_I|,|w^I|\le 5$. For the representative background (\ref{eq:buscher_G_and_B}), we enumerate the full truncated charge lattice, apply the Buscher charge map (\ref{eqn:buscher_charge_map}), and compare the exact original and dual tables. The Monte Carlo then uses $6\times 10^4$ independent sectors per ensemble and $2\times 10^3$ paired field draws with shared oscillators. Monodromy is measured on the stencil $\tau=0.14$ and $\sigma\in\{0,2\pi\}$. The oscillator/metric check uses $5\times 10^3$ shared oscillator draws at $(\tau,\sigma)=(0.23,1.11)$ and compares the empirical covariance, normalized by its trace, to the shapes of $G^{-1}$ and $\widetilde G^{-1}$. The dilaton is not sampled as a field; instead the code checks directly that the Buscher shift preserves $e^{-2\Phi}\sqrt{\det G}$.

\paragraph{Vertex operators:}
The operator-level tests use the chiral finite-mode sampler~\eref{eq:compact_chiral_field}--\eref{eq:compact_vertex_two_point} with $\alpha'=1$, $R=1.6$, $\widetilde R=0.625$, and $M=64$. One-point selection rules and the raw neutral/non-neutral two-point channels are measured with $1.2\times 10^5$ draws, at $(u,v)=(0.37,-0.41)$ for the one-point checks and at $(0,0)$, $(0.9,-0.9)$ for the raw two-point test. Normal-ordered equal-time scans use $2\times 10^5$ draws at the nine separations $\Delta\in[0.4,2.4]$, with insertions at $(0,0)$ and $(\Delta,-\Delta)$. The paired-draw T-duality identity uses $2048$ shared draws and $17$ random cylinder points, with $x_0\leftrightarrow\widetilde x_0$ modulo periods and $X_R\to -X_R$. Higher-charge two-point and mixed four-point correlators are evaluated with the blockwise covariance estimator described in the text: $2\times 10^5$ draws and $20$ blocks for the $(1,1)$ and $(2,1)$ two-point functions at $(0,0)$ and $(0.4,-0.4)$, and $10^5$ draws and $20$ blocks for the mixed four-point function at $(0,0)$, $(0.7,-0.3)$, $(1.4,-0.9)$, and $(2.1,-1.5)$.

\paragraph{Self-dual radius and enhanced current algebra:}
The self-dual checks reuse the finite-mode chiral conventions of the previous paragraph with $\alpha'=1$, $R=1$, $M=64$, and $2\times 10^5$ draws split into $20$ blocks. The Cartan current is represented numerically by the symmetric finite difference for $J^3_{L,\epsilon}$ used in the main text, with $\epsilon=\pi/M$. The Ward checks use the nine probe separations $\Delta\in[0.4,2.4]$ and the three-point points
\be
u_2=0.65 ~,\qquad u_3=2.40 ~,\qquad u_1\in\{1.15,1.40,1.65,1.90\} ~.
\ee
The non-Abelian current--primary channel $J_L^+g_{(-,+)}g_{(-,-)}$ uses the same $2\times 10^5$-draw, $20$-block setup, with $(u_2,v_2)=(0.9,-0.3)$, $(u_3,v_3)=(1.8,-0.9)$, and $u_1=u_2+\delta$ for seven values $\delta\in[0.08,0.30]$. The quoted coefficient is the inverse-variance weighted mean of the finite-$M$ normalized estimator $\hat c_+(\delta)$.

\paragraph{A toy T-fold from patchwise Buscher gluing:}
The patchwise construction reuses the same background as the Buscher test, with $\alpha'=1$, $\tau_2=0.5$, $\epsilon=0.5$, $\Lambda=6.0$, $N_\text{osc}=96$, and $|n_I|,|w^I|\le 5$. The base is approximated by $N_p=7$ patches at
\be
s_a=\frac{a}{N_p-1} ~,
\qquad a=0,\dots,N_p-1 ~,
\ee
with $G^{(a)}=(1-s_a)G+s_a\widetilde G$ and $B^{(a)}=(1-s_a)B+s_a\widetilde B$. A single set of oscillator latents is reused on all patches. The compact charge and zero-mode are sampled once in the initial chart and held fixed along the geometric part of the path, while the final overlap applies the Buscher charge map and returns the background from $(\widetilde G,\widetilde B)$ to $(G,B)$. Local covariance checks use $600$ oscillator draws per patch at $(\tau,\sigma)=(0.23,1.11)$. Seam-closure statistics use $250$ samples on the stencil $\tau=0$ and $\sigma\in\{0,\pi/2,\pi,3\pi/2\}$, recording both unwrapped RMS mismatch and wrapped RMS mismatch using angular differences. The generalized-metric analysis uses the same seven patch backgrounds, forms $\mathcal H(G,B)$ and $\Omega_y$ as in Section~\ref{sec:tdual_evidence}, and reports the endpoint mismatch before and after conjugation by the Buscher seam.

\bibliographystyle{JHEP}
\bibliography{refs}

@article{Capuozzo:2025ozt,
    author = "Capuozzo, Pietro and Robinson, Brandon and Suzzoni, Benjamin",
    title = "{Conformal Defects in Neural Network Field Theories}",
    eprint = "2512.07946",
    archivePrefix = "arXiv",
    primaryClass = "hep-th",
    month = "12",
    year = "2025"
}

@article{Demirtas:2023fir,
    author = "Demirtas, Mehmet and Halverson, James and Maiti, Anindita and Schwartz, Matthew D. and Stoner, Keegan",
    title = "{Neural network field theories: non-Gaussianity, actions, and locality}",
    eprint = "2307.03223",
    archivePrefix = "arXiv",
    primaryClass = "hep-th",
    doi = "10.1088/2632-2153/ad17d3",
    journal = "Mach. Learn. Sci. Tech.",
    volume = "5",
    number = "1",
    pages = "015002",
    year = "2024"
}

@article{Ferko:2025ogz,
    author = "Ferko, Christian and Halverson, James",
    title = "{Quantum Mechanics and Neural Networks}",
    eprint = "2504.05462",
    archivePrefix = "arXiv",
    primaryClass = "hep-th",
    month = "4",
    year = "2025"
}

@article{Osterwalder:1974tc,
	title        = {{Axioms for Euclidean Green's Functions. 2.}},
	author       = {Osterwalder, Konrad and Schrader, Robert},
	year         = 1975,
	journal      = {Commun. Math. Phys.},
	volume       = 42,
	pages        = 281,
	doi          = {10.1007/BF01608978},
	reportnumber = {Print-74-1480 (HARVARD)}
}

@article{Osterwalder:1973dx,
	title        = {{Axioms for Euclidean Green's functions}},
	author       = {Konrad Osterwalder and Robert Schrader},
	year         = 1973,
	journal      = {Communications in Mathematical Physics},
	publisher    = {Springer},
	volume       = 31,
	number       = 2,
	pages        = {83 -- 112}
}

@article{Ferko:2026axm,
    author = "Ferko, Christian and Halverson, James and Mutchler, Aaron",
    title = "{Universality of Neural Network Field Theory}",
    eprint = "2601.14453",
    archivePrefix = "arXiv",
    primaryClass = "hep-th",
    month = "1",
    year = "2026"
}

@article{Frank:2025zuk,
    author = "Frank, Samuel and Halverson, James and Maiti, Anindita and Ruehle, Fabian",
    title = "{Fermions and Supersymmetry in Neural Network Field Theories}",
    eprint = "2511.16741",
    archivePrefix = "arXiv",
    primaryClass = "hep-th",
    month = "11",
    year = "2025"
}

@misc{GarrigaAlonso2019DeepCN,
  author       = {Garriga{-}Alonso, Adri{\`a} and Aitchison, Laurence and Rasmussen, Carl Edward},
  title        = {Deep Convolutional Networks as Shallow Gaussian Processes},
  year         = {2019},
  eprint       = {1808.05587},
  archivePrefix= {arXiv},
  primaryClass = {cs.LG}
}

@article{Halverson:2020trp,
    author = "Halverson, James and Maiti, Anindita and Stoner, Keegan",
    title = "{Neural Networks and Quantum Field Theory}",
    eprint = "2008.08601",
    archivePrefix = "arXiv",
    primaryClass = "cs.LG",
    doi = "10.1088/2632-2153/abeca3",
    journal = "Mach. Learn. Sci. Tech.",
    volume = "2",
    number = "3",
    pages = "035002",
    year = "2021"
}

@article{Halverson:2021aot,
    author = "Halverson, James",
    title = "{Building Quantum Field Theories Out of Neurons}",
    eprint = "2112.04527",
    archivePrefix = "arXiv",
    primaryClass = "hep-th",
    month = "12",
    year = "2021"
}

@article{Halverson:2024axc,
    author = "Halverson, James and Naskar, Joydeep and Tian, Jiahua",
    title = "{Conformal fields from neural networks}",
    eprint = "2409.12222",
    archivePrefix = "arXiv",
    primaryClass = "hep-th",
    doi = "10.1007/JHEP10(2025)039",
    journal = "JHEP",
    volume = "10",
    pages = "039",
    year = "2025"
}

@article{Huang:2025ipy,
    author = "Huang, Guojun and Zhou, Kai",
    title = "{The neural networks with tensor weights and emergent fermionic Wick rules in the large-width limit}",
    eprint = "2507.05303",
    archivePrefix = "arXiv",
    primaryClass = "hep-th",
    doi = "10.1016/j.physletb.2025.140146",
    journal = "Phys. Lett. B",
    volume = "873",
    pages = "140146",
    year = "2026"
}

@article{Sen:2025vzl,
    author = "Sen, Srimoyee and Vaidya, Varun",
    title = "{Viability of perturbative expansion for quantum field theories on neurons}",
    eprint = "2508.03810",
    archivePrefix = "arXiv",
    primaryClass = "hep-th",
    month = "8",
    year = "2025"
}

@article{StromingerYauZaslow1996,
  author        = {Strominger, Andrew and Yau, Shing-Tung and Zaslow, Eric},
  title         = {Mirror Symmetry is T-Duality},
  eprint        = {hep-th/9606040},
  archivePrefix = {arXiv},
  journal       = {Nucl. Phys. B},
  volume        = {479},
  number        = {1-2},
  pages         = {243--259},
  year          = {1996},
  doi           = {10.1016/0550-3213(96)00434-8}
}

@article{HoriVafa2000,
  author        = {Hori, Kentaro and Vafa, Cumrun},
  title         = {Mirror Symmetry},
  eprint        = {hep-th/0002222},
  archivePrefix = {arXiv},
  reportNumber  = {HUTP-00-A005},
  year          = {2000}
}

@article{Halverson:2024hax,
    author = "Halverson, Jim",
    title = "{TASI Lectures on Physics for Machine Learning}",
    eprint = "2408.00082",
    archivePrefix = "arXiv",
    primaryClass = "hep-th",
    month = "7",
    year = "2024"
}

@misc{hron2020infinite,
  author       = {Hron, Jakub and Bahri, Yasaman and Sohl{-}Dickstein, Jascha and Novak, Roman},
  title        = {Infinite Attention: {NNGP} and {NTK} for Deep Attention Networks},
  year         = {2020},
  eprint       = {2006.10540},
  archivePrefix= {arXiv},
  primaryClass = {stat.ML}
}

@inproceedings{Jacot2018NeuralTK,
  author    = {Jacot, Arthur and Gabriel, Franck and Hongler, Cl{\'e}ment},
  title     = {Neural Tangent Kernel: Convergence and Generalization in Neural Networks},
  booktitle = {Advances in Neural Information Processing Systems},
  year      = {2018}
}

@article{Maiti:2021fpy,
    author = "Maiti, Anindita and Stoner, Keegan and Halverson, James",
    title = "{Symmetry-via-Duality: Invariant Neural Network Densities from Parameter-Space Correlators}",
    eprint = "2106.00694",
    archivePrefix = "arXiv",
    primaryClass = "cs.LG",
    month = "6",
    year = "2021"
}

@misc{Matthews2018GaussianPB,
  author       = {Matthews, Alexander G. de G. and Rowland, Mark and Hron, Jakub and Turner, Richard E. and Ghahramani, Zoubin},
  title        = {Gaussian Process Behaviour in Wide Deep Neural Networks},
  year         = {2018},
  eprint       = {1804.11271},
  archivePrefix= {arXiv},
  primaryClass = {stat.ML}
}

@phdthesis{neal,
	title        = {BAYESIAN LEARNING FOR NEURAL NETWORKS},
	author       = {Neal, Radford M.},
	year         = 1995,
	school       = {University of Toronto}
}

@misc{Novak2018BayesianCN,
  author       = {Novak, Roman and Xiao, Lechao and Lee, Jaehoon and Bahri, Yasaman and Abolafia, Daniel A. and Pennington, Jeffrey and Sohl{-}Dickstein, Jascha},
  title        = {Bayesian Convolutional Neural Networks with Many Channels are Gaussian Processes},
  year         = {2018},
  eprint       = {1810.05148},
  archivePrefix= {arXiv},
  primaryClass = {cs.LG}
}

@article{PhysRevE.104.064301,
  author  = {Naveh, Guy and Ben David, Omri and Sompolinsky, Haim and Ringel, Zohar},
  title   = {Predicting the Outputs of Finite Deep Neural Networks Trained with Noisy Gradients},
  journal = {Phys. Rev. E},
  volume  = {104},
  pages   = {064301},
  year    = {2021},
  doi     = {10.1103/PhysRevE.104.064301}
}

@inproceedings{rahimi2007random,
  author    = {Rahimi, Ali and Recht, Benjamin},
  title     = {Random Features for Large-Scale Kernel Machines},
  booktitle = {Advances in Neural Information Processing Systems},
  volume    = {20},
  year      = {2007}
}

@article{Robinson:2025ybg,
    author = "Robinson, Brandon",
    title = "{Virasoro Symmetry in Neural Network Field Theories}",
    eprint = "2512.24420",
    archivePrefix = "arXiv",
    primaryClass = "hep-th",
    month = "12",
    year = "2025"
}

@book{roberts2022principles,
  author    = {Roberts, Daniel A. and Yaida, Sho and Hanin, Boris},
  title     = {The Principles of Deep Learning Theory},
  volume    = {46},
  publisher = {Cambridge University Press},
  address   = {Cambridge, MA, USA},
  year      = {2022}
}

@misc{schoenholz2017correspondence,
  author       = {Schoenholz, Samuel S. and Pennington, Jeffrey and Sohl{-}Dickstein, Jascha},
  title        = {A Correspondence between Random Neural Networks and Statistical Field Theory},
  year         = {2017},
  eprint       = {1710.06570},
  archivePrefix= {arXiv},
  primaryClass = {stat.ML}
}

@inproceedings{vaswani2017attention,
  author    = {Vaswani, Ashish and Shazeer, Noam and Parmar, Niki and Uszkoreit, Jakob and Jones, Llion and Gomez, Aidan N. and Kaiser, {\L}ukasz and Polosukhin, Illia},
  title     = {Attention Is All You Need},
  booktitle = {Advances in Neural Information Processing Systems},
  volume    = {30},
  year      = {2017}
}

@inproceedings{williams,
  author    = {Williams, Christopher K. I.},
  title     = {Computing with Infinite Networks},
  booktitle = {Advances in Neural Information Processing Systems},
  volume    = {9},
  editor    = {Mozer, Michael and Jordan, Michael and Petsche, Thomas},
  publisher = {MIT Press},
  year      = {1996},
  url       = {https://proceedings.neurips.cc/paper\_files/paper/1996/file/ae5e3ce40e0404a45ecacaaf05e5f735-Paper.pdf}
}

@misc{Yaida2019NonGaussianPA,
  author       = {Yaida, Sho},
  title        = {Non-Gaussian Processes and Neural Networks at Finite Widths},
  year         = {2019},
  eprint       = {1910.00019},
  archivePrefix= {arXiv},
  primaryClass = {cs.LG}
}

@misc{yangTP1,
  author       = {Yang, Greg},
  title        = {Tensor Programs {I}: Wide Feedforward or Recurrent Neural Networks of Any Architecture are Gaussian Processes},
  year         = {2019},
  eprint       = {1910.12478},
  archivePrefix= {arXiv},
  primaryClass = {cs.NE}
}

@misc{yangTP2,
  author       = {Yang, Greg},
  title        = {Tensor Programs {II}: Neural Tangent Kernel for Any Architecture},
  year         = {2020},
  eprint       = {2006.14548},
  archivePrefix= {arXiv},
  primaryClass = {cs.LG}
}

@article{Frank:2026bui,
    author = "Frank, Samuel and Halverson, James",
    title = "{String Theory from Infinite Width Neural Networks}",
    eprint = "2601.06249",
    archivePrefix = "arXiv",
    primaryClass = "hep-th",
    month = "1",
    year = "2026"
}

@article{Berezinskii1971,
  author  = {Berezinskii, V. L.},
  title   = {Destruction of Long-range Order in One-dimensional and Two-dimensional Systems Having a Continuous Symmetry Group I. Classical Systems},
  journal = {Sov. Phys. JETP},
  volume  = {32},
  pages   = {493--500},
  year    = {1971}
}

@article{KosterlitzThouless1973,
  author  = {Kosterlitz, J. M. and Thouless, D. J.},
  title   = {Ordering, metastability and phase transitions in two-dimensional systems},
  journal = {J. Phys. C},
  volume  = {6},
  pages   = {1181--1203},
  year    = {1973}
}

@article{Kosterlitz1974,
  author  = {Kosterlitz, J. M.},
  title   = {The critical properties of the two-dimensional {XY} model},
  journal = {J. Phys. C},
  volume  = {7},
  pages   = {1046--1060},
  year    = {1974}
}

@article{Jose1977,
  author  = {Jos{\'e}, J. V. and Kadanoff, L. P. and Kirkpatrick, S. and Nelson, D. R.},
  title   = {Renormalization, vortices, and symmetry-breaking perturbations in the two-dimensional planar model},
  journal = {Phys. Rev. B},
  volume  = {16},
  pages   = {1217--1241},
  year    = {1977}
}

@article{MerminWagner1966,
  author  = {Mermin, N. D. and Wagner, H.},
  title   = {Absence of Ferromagnetism or Antiferromagnetism in One- or Two-Dimensional Isotropic Heisenberg Models},
  journal = {Phys. Rev. Lett.},
  volume  = {17},
  pages   = {1133--1136},
  year    = {1966}
}

@article{Minnhagen1987,
  author  = {Minnhagen, Petter},
  title   = {The two-dimensional Coulomb gas, vortex unbinding, and superfluid-superconducting films},
  journal = {Rev. Mod. Phys.},
  volume  = {59},
  pages   = {1001--1066},
  year    = {1987}
}

@article{bishop1978study,
  title={Study of the superfluid transition in two-dimensional He 4 films},
  author={Bishop, DJ and Reppy, JD},
  journal={Physical Review Letters},
  volume={40},
  number={26},
  pages={1727},
  year={1978},
  publisher={APS}
}

@article{Buscher:1987qj,
    author = "Buscher, T. H.",
    title = "{Path Integral Derivation of Quantum Duality in Nonlinear Sigma Models}",
    reportNumber = "ITP-SB-87-61",
    doi = "10.1016/0370-2693(88)90602-8",
    journal = "Phys. Lett. B",
    volume = "201",
    pages = "466--472",
    year = "1988"
}

@article{Buscher:1987sk,
    author = "Buscher, T. H.",
    title = "{A Symmetry of the String Background Field Equations}",
    reportNumber = "ITP-SB-87-21",
    doi = "10.1016/0370-2693(87)90769-6",
    journal = "Phys. Lett. B",
    volume = "194",
    pages = "59--62",
    year = "1987"
}

@article{Hull:2006qs,
    author = "Hull, C. M.",
    title = "{Global aspects of T-duality, gauged sigma models and T-folds}",
    eprint = "hep-th/0604178",
    archivePrefix = "arXiv",
    reportNumber = "IMPERIAL-TP-06-CH-01",
    doi = "10.1088/1126-6708/2007/10/057",
    journal = "JHEP",
    volume = "10",
    pages = "057",
    year = "2007"
}

@article{Hull:2006va,
    author = "Hull, C M",
    title = "{Doubled Geometry and T-Folds}",
    eprint = "hep-th/0605149",
    archivePrefix = "arXiv",
    reportNumber = "IMPERIAL-TP-06-CH-02",
    doi = "10.1088/1126-6708/2007/07/080",
    journal = "JHEP",
    volume = "07",
    pages = "080",
    year = "2007"
}

@article{Hull:2004in,
    author = "Hull, C. M.",
    title = "{A Geometry for non-geometric string backgrounds}",
    eprint = "hep-th/0406102",
    archivePrefix = "arXiv",
    reportNumber = "IMPERIAL-TP-3-04-13",
    doi = "10.1088/1126-6708/2005/10/065",
    journal = "JHEP",
    volume = "10",
    pages = "065",
    year = "2005"
}

@article{Hellerman:2002ax,
    author = "Hellerman, Simeon and McGreevy, John and Williams, Brook",
    title = "{Geometric constructions of nongeometric string theories}",
    eprint = "hep-th/0208174",
    archivePrefix = "arXiv",
    reportNumber = "SLAC-PUB-9510, SU-ITP-02-35",
    doi = "10.1088/1126-6708/2004/01/024",
    journal = "JHEP",
    volume = "01",
    pages = "024",
    year = "2004"
}

@article{Dabholkar:2002sy,
    author = "Dabholkar, Atish and Hull, Chris",
    title = "{Duality twists, orbifolds, and fluxes}",
    eprint = "hep-th/0210209",
    archivePrefix = "arXiv",
    reportNumber = "TIFR-TH-01-24, QMUL-PH-02-16",
    doi = "10.1088/1126-6708/2003/09/054",
    journal = "JHEP",
    volume = "09",
    pages = "054",
    year = "2003"
}

@article{Gross:1985fr,
    author = "Gross, David J. and Harvey, Jeffrey A. and Martinec, Emil J. and Rohm, Ryan",
    title = "{Heterotic String Theory. 1. The Free Heterotic String}",
    reportNumber = "PRINT-85-0203 (PRINCETON)",
    doi = "10.1016/0550-3213(85)90394-3",
    journal = "Nucl. Phys. B",
    volume = "256",
    pages = "253",
    year = "1985"
}

@article{Narain:1985jj,
    author = "Narain, K. S.",
    title = "{New Heterotic String Theories in Uncompactified Dimensions {\ensuremath{<}} 10}",
    reportNumber = "RAL-85-097",
    doi = "10.1016/0370-2693(86)90682-9",
    journal = "Phys. Lett. B",
    volume = "169",
    pages = "41--46",
    year = "1986"
}

@article{Witten:1983ar,
    author = "Witten, Edward",
    editor = "Stone, M.",
    title = "{Nonabelian Bosonization in Two-Dimensions}",
    reportNumber = "PRINT-83-0934 (PRINCETON)",
    doi = "10.1007/BF01215276",
    journal = "Commun. Math. Phys.",
    volume = "92",
    pages = "455--472",
    year = "1984"
}

@book{Polchinski:1998rq,
    author = "Polchinski, J.",
    title = "{String theory. Vol. 1: An introduction to the bosonic string}",
    doi = "10.1017/CBO9780511816079",
    isbn = "978-0-511-25227-3, 978-0-521-67227-6, 978-0-521-63303-1",
    publisher = "Cambridge University Press",
    series = "Cambridge Monographs on Mathematical Physics",
    month = "12",
    year = "2007"
}

\end{document}